\documentclass[journal,11pt,draftclsnofoot,onecolumn]{IEEEtran}

\usepackage{graphicx}
\usepackage[square,comma,numbers,sort&compress]{natbib}
\usepackage{amsthm,amssymb,amscd,amssymb,amsmath,mathrsfs}
\usepackage{balance}
\usepackage[dvipdfmx,colorlinks,linkcolor=blue,anchorcolor=blue,citecolor=red]{hyperref}  
\usepackage{xcolor}
\usepackage{overpic}
\makeatletter
\newcommand{\rmnum}[1]{\romannumeral #1}
\newcommand{\Rmnum}[1]{\expandafter\@slowromancap\romannumeral #1@}
\makeatother

\theoremstyle{plain}

\newtheorem{pro}{Proposition}
\newtheorem{lem}{Lemma}

\theoremstyle{definition}
\newtheorem{defn}{Definition}

\theoremstyle{remark}
\newtheorem{remark}{Remark}

\begin{document}
%
\title{Delay-Robustness in Distributed Control of Timed Discrete-Event Systems Based on Supervisor Localization\\
 \Large(\today)}

\author{Renyuan Zhang, Kai Cai, Yongmei Gan, W.M. Wonham
\thanks{R. Zhang is with School of Automation, Northwestern Polytechnical University, China; K. Cai is with Urban Research Plaza, Osaka City University, Japan; Y. Gan is with School of Electrical Engineering, Xi¡¯an Jiaotong University, China; and W.M. Wonham is with the Systems Control Group, Department of Electrical and Computer Engineering, University of Toronto,
Canada. (Emails: ryzhang@nwpu.edu.cn;
kai.cai@info.eng.osaka-cu.ac.jp; ymgan@mail.xjtu.edu.cn;
wonham@control.utoronto.ca). } }

\maketitle

\thispagestyle{empty} \pagestyle{plain}

\begin{abstract}
Recently we studied communication delay in distributed control of
untimed discrete-event systems based on supervisor localization.
We proposed a property called delay-robustness: the overall system
behavior controlled by distributed controllers with communication
delay is logically equivalent to its delay-free counterpart. In this paper we
extend our previous work to timed discrete-event systems, in which
communication delays are counted by a special clock event {\it tick}.
First, we propose a timed channel model and define timed
delay-robustness; for the latter, a polynomial verification procedure
is presented. Next, if the delay-robust property does not hold, we
introduce \emph{bounded} delay-robustness, and present an algorithm
to compute the \emph{maximal} delay bound (measured by number of $tick$s)
for transmitting a channeled event. Finally, we demonstrate delay-robustness
on the example of an under-load tap-changing transformer.

\end{abstract}


\section{Introduction} \label{Sec:Intro}


For distributed control of discrete-event systems (DES), supervisor
localization was recently proposed \cite{CaiWonham:2010a,
CaiWonham:2010b,ZhangCai:2013,CaiZhangWonham:2013} which
decomposes a monolithic
supervisor or a heterarchical array of modular supervisors into
local controllers for individual agents.  Collective local
controlled behavior is guaranteed to be globally optimal and
nonblocking, assuming that the shared events among local controllers
are communicated instantaneously, i.e. with no delay.  In practice,
however, local controllers are linked by a physical communication
network in which delays may be inevitable. Hence, for correct
implementation of the local controllers obtained by localization, it
is essential to model and appraise communication delays.

In \cite{ZhangCai:arXiv2012} and its conference precursor
\cite{ZhangCai:conf12}, we studied communication delays among local controllers for untimed
DES.  In particular, we proposed a new concept called
\emph{delay-robustness}, meaning that the systemic behavior of local
controllers interconnected by communication channels subject to
unbounded delays is logically equivalent to its delay-free counterpart.
Moreover, we designed an efficient procedure to verify for which
channeled events the system is delay-robust.  If for a channeled
event $r$ the system fails to be delay-robust, there may still exist a
finite bound for which the system can tolerate a delay in $r$. In unitimed DES,
however, there lacks a \emph{temporal} measure for the delay bound (except for counting the
number of occurrences of untimed events).

In this paper and its conference antecedent\cite{ZhangCaiWonham:2014}, we extend our study on delay-robustness to the timed
DES (or TDES) framework proposed by Brandin and Wonham
\cite{BrandinWonham:94,Wonham:2013a}. In this framework, the special
clock event \emph{tick} provides a natural way of modeling
communication delay as temporal behavior. We first propose a timed
channel model for transmitting each channeled event, which
effectively measures communication delay by the number of \emph{tick}
occurrences, with no {\it a priori} upper bound, so that the channel models
\emph{unbounded} delay.  We then define timed delay-robustness with
respect to the timed channel, thus extending its untimed counterpart
\cite{ZhangCai:arXiv2012,ZhangCai:conf12} in two respects: (1)
the system's temporal behavior is accounted for, and (2) timed
controllability is required. A polynomial algorithm is presented to
verify timed delay-robustness according to this new definition.

If the delay-robust property fails to hold, we introduce
\emph{bounded} delay-robustness and present a corresponding
verification algorithm. In particular, the
algorithm computes the \emph{maximal} delay bound (in terms of
number of $tick$s) for transmitting a channeled event, i.e. the
largest delay that can be tolerated without violating the system
specifications.  These concepts and the
corresponding algorithms are illustrated for the case of an
under-load tap-changing transformer (ULTC).

Distributed/decentralized supervisory control with communication
delay has been widely studied for untimed DES (e.g.
\cite{BarrettLafortune:2000, Tripakis:2004, Schmidt:2007, Xu:2008,
Hiraishi:2009, Kalyon:2011, Lin:2014, Darondeau:2012, SadidRicker:2014}). In
particular in \cite{Tripakis:2004,Kalyon:2011}, the existence of
distributed controllers in the unbounded delay case is proved to be
undecidable; and in \cite{Tripakis:2004, Schmidt:2007, Xu:2008,
Hiraishi:2009, Lin:2014}, distributed controllers are synthesized
under the condition that communication delay is bounded. We also note that Sadid et al. \cite{SadidRicker:2014} propose a way to verify robustness of a given synchronous protocol with respect to a fixed or a finitely-bounded delay, as measured by the number of untimed events occurring during  the transmitting process. We refer to
\cite{ZhangCai:arXiv2012,ZhangCai:conf12} for a detailed review
of these works and their differences from our approach.
Communication delay in timed DES, on the other hand, has (to our
knowledge) received little attention.  The present work is based on our
previous research on timed supervisor localization
\cite{ZhangCai:2013,CaiZhangWonham:2013}.

The paper is organized as follows. Sect.~\ref{Sec:BackGrouMotiva}
provides a review of the Brandin-Wonham TDES framework and
recalls supervisor localization for TDES.  In
Sect.~\ref{Sec:DelayRobust} we introduce a timed channel model, and
present the concept and verification algorithm for timed
delay-robustness.In Sect.~\ref{Sec:MaxDelayBound} we define
bounded delay-robustness, and present an algorithm to compute the
maximal delay bound. These concepts and the
corresponding algorithms are demonstrated in
Sect.~\ref{Sec:DSCofULTC} on the distributed control problem for an under-load
tap-changing transformer (ULTC) with communications. Conclusions
are presented in Sect.~\ref{Sec:Concl}.

\section{Distributed Control by Supervisor Localization of TDES} \label{Sec:BackGrouMotiva}

\subsection{Preliminaries on TDES} \label{SubSec:Prelimin}


The TDES model proposed by Brandin and Wonham
\cite{BrandinWonham:94} is and extension of the untimed DES generator model
of the Ramadge-Wonham framework \cite{Wonham:2013a}. A TDES is given
by
\begin{equation} \label{eq:TDES}
{\bf G} := (Q, \Sigma, \delta, q_0, Q_m).
\end{equation}
Here $Q$ is the finite set of \emph{states}; $\Sigma$ is the finite
set of events including the special event $tick$, which represents
``tick of the global clock''; $\delta:Q\times\Sigma \rightarrow Q$
is the (partial) {\it state transition function} (this is derived
from the corresponding activity transition function; the reader is
referred to the detailed transition rules given in
\cite{BrandinWonham:94,Wonham:2013a}); $q_0$ is the {\it initial
state}; and $Q_m \subseteq Q$ is the set of {\it marker states}. The
transition function is extended to $\delta:Q\times \Sigma^*
\rightarrow Q$ in the usual way. The {\it closed behavior} of $\bf
G$ is the language $L({\bf G}) := \{s \in \Sigma^*|\delta(q_0,s)!\}$
and the {\it marked behavior} is $L_m({\bf G}) := \{s \in L({\bf
G})| \delta(q_0, s) \in Q_m\} \subseteq L({\bf G})$. We say that
$\bf G$ is \emph{nonblocking} if $\bar{L}_m({\bf G}) = L({\bf G})$,
where $\bar{\cdot}$ denotes {\it prefix closure}
\cite{Wonham:2013a}.

Let $\Sigma^*$ be the set of all finite strings, including the empty
string $\epsilon$. For $\Sigma' \subseteq \Sigma$, the \emph{natural
projection} $P : \Sigma^* \rightarrow \Sigma'^*$ is defined by
\begin{equation} \label{eq:natpro}
\begin{split}
P(\epsilon) &= \epsilon; \\
P(\sigma) &= \left\{
  \begin{array}{ll}
    \epsilon, & \hbox{if $\sigma \notin \Sigma'$,} \\
    \sigma, & \hbox{if $\sigma \in \Sigma'$;}
  \end{array}
\right.\\
P(s\sigma) &= P(s)P(\sigma),\ \ s \in \Sigma^*, \sigma \in \Sigma.
\end{split}
\end{equation}
As usual, $P$ is extended to $P : Pwr(\Sigma^*) \rightarrow
Pwr(\Sigma'^*)$, where $Pwr(\cdot)$ denotes powerset. Write $P^{-1}
: Pwr(\Sigma'^*) \rightarrow Pwr(\Sigma^*)$ for the
\emph{inverse-image function} of $P$.

To adapt the TDES $\bf G$ in (\ref{eq:TDES}) for supervisory control, we
first designate a subset of events, denoted by $\Sigma_{hib}
\subseteq \Sigma$, to be the {\it prohibitible} events which can be
disabled by an external supervisor. Next, and specific to TDES, we
bring in another category of events, called the {\it forcible}
events, which can \emph{preempt} event $tick$; let $\Sigma_{for}
\subseteq \Sigma$ denote the set of forcible events. Note that $tick
\notin \Sigma_{hib} \cup \Sigma_{for}$. Now it is convenient to
define the {\it controllable} event set $\Sigma_c :=
\Sigma_{hib}~\dot\cup~\{tick\}$. The {\it uncontrollable} event set
is $\Sigma_{u} := \Sigma - \Sigma_c$.

We introduce the notion of (timed) controllability as follows. For a
string $s \in L({\bf G})$, define $Elig_{\bf G}(s):=\{\sigma \in
\Sigma|s\sigma \in L({\bf G})\}$ to be the subset of events
`eligible' to occur (i.e. defined) at the state $q = \delta(q_0,
s)$. Consider an arbitrary language $F \subseteq L({\bf G})$ and a
string $s \in \overline{F}$; similarly define the eligible event
subset $Elig_F(s):= \{\sigma \in \Sigma|s\sigma \in \overline{F}\}$.
We say $F$ is {\it controllable} with respect to $\bf G$ if, for all
$s \in \overline{F}$,
\begin{eqnarray} \label{eq:defcontrol}
Elig_F(s)\supseteq
\left\{
   \begin{array}{l}
      Elig_{\bf G}(s)\cap(\Sigma_u \dot{\cup}\{tick\})\\
      ~~~~~~~~~~~~~~~\text{if}~~Elig_F(s)\cap \Sigma_{for} = \emptyset,\\
      Elig_{\bf G}(s)\cap \Sigma_u \\
      ~~~~~~~~~~~~~~~\text{if}~~Elig_F(s)\cap \Sigma_{for} \neq \emptyset.
   \end{array}
\right.
\end{eqnarray}
Whether or not $F$ is controllable, we denote by $\mathcal{C}(F)$ the set of all controllable sublanguages of $F$. Then $\mathcal{C}(F)$ is nonempty, closed under arbitrary set unions, and thus contains a unique supremal (largest) element denoted by $sup\mathcal{C}(F)$ \cite{BrandinWonham:94,Wonham:2013a}. Now consider a specification language $E \subseteq \Sigma^*$ imposed on the timed behavior of
$\bf G$; $E$ may represent a logical and/or temporal requirement. Let the TDES
\begin{equation}\label{eq:MonoliSUP}
{\bf SUP} = (X, \Sigma, \xi, x_0, X_m)
\end{equation}
be the corresponding \emph{monolithic supervisor} that is optimal (i.e., maximally permissive) and nonblocking in the following sense: $\bf SUP$'s marked language $L_m({\bf SUP})$ is
\begin{equation*}
L_m({\bf SUP}) = sup\mathcal{C}(E\cap L_m({\bf G})) \subseteq L_m(\bf G)
\end{equation*}
and moreover its closed language $L({\bf SUP})$ is $L({\bf SUP}) = \overline{L}_m({\bf SUP}).$

\subsection{Supervisor Localization of TDES} \label{SubSec:SuperviLocali}


In this subsection, we introduce the supervisor localization
procedure, which was initially proposed in the untimed DES framework
\cite{CaiWonham:2010a} and then adapted to the TDES framework
\cite{ZhangCai:2013,CaiZhangWonham:2013}. By this procedure, a set of \emph{local
controllers} and \emph{local preemptors} is obtained and shown to be
`control equivalent' to the monolithic supervisor $\bf SUP$ in
(\ref{eq:MonoliSUP}). By allocating these constructed local
controllers and preemptors to each component agent, we build a
distributed supervisory control architecture.

Let TDES $\bf G$ in (\ref{eq:TDES}) be the plant to be controlled
and $E$ be a specification language. As in \cite{Wonham:2013a},
synthesize the monolithic optimal and nonblocking supervisor ${\bf
SUP}$. Supervisor $\bf SUP$'s control action includes (i) disabling
prohibitible events in $\Sigma_{hib}$ and (ii) preempting $tick$ via
forcible events in $\Sigma_{for}$. By the supervisor localization
procedure, a set of local controllers $\{{\bf LOC}_\alpha^C
\mbox{defined on } \Sigma_\alpha | \alpha \in \Sigma_{hib}\}$ and a
set of local preemptors $\{{\bf LOC}_\beta^P \mbox{defined on }
\Sigma_\beta | \beta \in \Sigma_{for}\}$ are constructed. These
${\bf LOC}_\alpha^C$ and ${\bf LOC}_\beta^P$ are all TDES as in
(\ref{eq:TDES}), and proved to be control equivalent to ${\bf SUP}$
(with respect to ${\bf G}$) in the following sense:
\begin{align}
   L({\bf G}) \cap &\Big(\mathop \bigcap\limits_{\alpha \in \Sigma_{hib}}P_\alpha^{-1}L({\bf LOC}^C_{\alpha}) \Big)
   \cap\Big(\mathop \bigcap\limits_{\beta \in \Sigma_{for}}P_\beta^{-1}L({\bf LOC}^P_{\beta}) \Big) = L({\bf SUP}), \label{eq:sub1:ControlEqu}\\
   L_m({\bf G}) \cap &\Big(\mathop \bigcap\limits_{\alpha \in \Sigma_{hib}}P_\alpha^{-1}L_m({\bf LOC}^C_{\alpha}) \Big)
   \cap\Big(\mathop \bigcap\limits_{\beta \in \Sigma_{for}}P_\beta^{-1}L_m({\bf LOC}^P_{\beta}) \Big) = L_m({\bf SUP}). \label{eq:sub2:ControlEqu}
\end{align}
Here $P_{\alpha} : \Sigma^* \rightarrow \Sigma_{\alpha}^*$ and $P_{\beta} : \Sigma^* \rightarrow \Sigma_{\beta}^*$ are the natural projections as in (\ref{eq:natpro}).

Now, using the constructed local controllers and local preemptors,
we build a distributed supervisory control architecture (without
communication delay) for a multi-agent TDES plant. Consider that the
plant $\bf G$ consists of $N$ component TDES ${\bf G}_i$ ($i \in
\mathcal {N} := \{1,2,...,N\}$), each with event set $\Sigma_i \ni
tick$. For simplicity assume $\Sigma_{i} \cap \Sigma_{j} =
\{tick\}$, for all $i \neq j \in \mathcal {N}$; namely the agents
${\bf G}_i$ are independent except for synchronization on the global
event $tick$. As a result, the marked and closed behaviors of the composition
of ${\bf G}_i$ coincide with those of their synchronous product
\cite{Wonham:2013a}, and thus we use synchronous product instead of
composition to combine TDES together, i.e. ${\bf G} =
\mathop{||}\limits_{i\in \mathcal{N}}{\bf G}_i$ where $||$ denotes the
synchronous product of TDES.\footnote{The closed and marked
behaviors of ${\bf TDES} = {\bf TDES1}\mathop{||}{\bf TDES2}$ are
$L({\bf TDES}) = L({\bf TDES1})\mathop{||} L({\bf TDES2})$ and
$L_m({\bf TDES}) = L_m({\bf TDES1}) \mathop{||} L_m({\bf TDES2})$,
where $||$ denotes the synchronous product of languages
\cite{Wonham:2013a}.}

A convenient {\it allocation policy} of local controllers/preemptors
is the following. For a fixed agent ${\bf G}_i$, let
$\Sigma_{i,for}, \Sigma_{i,hib} \subseteq \Sigma_i$ be its forcible
event set and prohibitible event set, respectively. Then allocate to
${\bf G}_i$ the set of local controllers ${\bf LOC}_{i}^C:=\{{\bf
LOC}_{\alpha}^C|\alpha \in \Sigma_{i,hib}\}$ and the set of local
preemptors ${\bf LOC}_{i}^P:=\{{\bf LOC}_{\beta}^P|\beta \in
\Sigma_{i,for}\}$. This allocation creates a distributed control
architecture for the multi-agent plant ${\bf G}$, in which each
agent ${\bf G}_i$ is controlled by its own local
controllers/preemptors, while interacting with other agents through
communication of shared events. For agent ${\bf G}_i$, the set of
{\it communication events} that need to be imported from other
agents is
\begin{align} \label{eq:ComEvent}
   \Sigma_{com,i} := \Big(\mathop \bigcup\limits_{\alpha \in \Sigma_{i,hib}}\Sigma_\alpha - \Sigma_i \Big)
   \cup \Big(\mathop \bigcup\limits_{\beta \in \Sigma_{i, for}}\Sigma_\beta - \Sigma_i \Big)
\end{align}
where $\Sigma_\alpha$ and $\Sigma_\beta$ are the event sets of ${\bf LOC}_{\alpha}^C$ and of ${\bf LOC}_{\beta}^P$ respectively.

However, this distributed control architecture is built under the
assumption that the communication delay of communication events is
negligible. While simplifying the design of distributed controllers,
this assumption may be unrealistic in practice, where controllers
are linked by a physical network subject to delay. In the rest of
this paper, we investigate how the communication delay affects the
synthesized local control strategies and the corresponding overall
system behavior.

\section{Timed Delay-Robustness}\label{Sec:DelayRobust}



Consider event communication between a pair of agents ${\bf G}_i$
and ${\bf G}_j$ ($i,j \in \mathcal {N}$): specifically, ${\bf G}_j$
sends an event $\sigma$ to ${\bf G}_i$.  Let $\Sigma_j$ be the event
set of ${\bf G}_j$ and $\Sigma_{com,i}$ as in (\ref{eq:ComEvent})
the set of communication events that ${\bf G}_i$ imports from other
agents. Then the set of events that ${\bf G}_j$ sends to ${\bf G}_i$
is
\begin{align}\label{eq:ChnEvent}
\Sigma_{j,com,i} := \Sigma_j \cap \Sigma_{com,i}.
\end{align}
We thus have event $\sigma \in \Sigma_{j,com,i}$.

\begin{figure}[!t]
\centering
    \includegraphics[scale = 0.6]{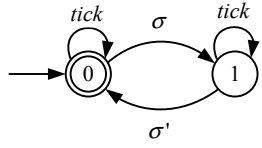}
\caption{Timed channel model ${\bf CH}({j,\sigma,i})$ for
transmitting event $\sigma$ from ${\bf G}_j$ to ${\bf G}_i$ with
indefinite (i.e. unspecified) time delay.} \label{fig:ComChannel}
\end{figure}

Now consider the timed channel model ${\bf CH}(j,\sigma,i)$ for
$\sigma$ transmission displayed in Fig.~\ref{fig:ComChannel}. ${\bf
CH}(j,\sigma,i)$ is a 2-state TDES with event set $\{\sigma,
\sigma', tick\}$. The transition from state 0 to 1 by $\sigma$ means
that ${\bf G}_j$ has sent $\sigma$ to channel, while the transition
from state 1 back to 0 by $\sigma'$ means that ${\bf G}_i$ has
received $\sigma$ from channel. We refer to $\sigma'$ as the
\emph{signal event} of $\sigma$, and assign its controllability
status to be the same as $\sigma$ (i.e. $\sigma'$ is controllable iff
$\sigma$ is controllable). The selfloop transition $tick$ at state 1
therefore counts communication delay of $\sigma$ transmission: the
number of $tick$s that elapses between $\sigma$ and $\sigma'$.
Measuring delay by $tick$ events is a major improvement compared to
the untimed channel model we used in \cite{ZhangCai:conf12} where no
suitable measure exists to count delay. Later in
Sect.~\ref{Sec:MaxDelayBound}, with the aid of this measure we will
compute useful delay bounds for event communication.

It should be stressed that the number of $tick$ occurrences between
$\sigma$ and $\sigma'$ is unspecified, inasmuch as the selfloop
$tick$ at state 1 may occur indefinitely. In this sense, ${\bf
CH}(j,\sigma,i)$ models possibly \emph{unbounded} communication
delay. Note that $tick$ is also selflooped at state 0; this is not
used to count delay, but rather for the technical necessity of preventing the
event $tick$ from being blocked when synchronizing ${\bf CH}(j,\sigma,i)$
with other TDES. The initial state 0 is marked, signaling each
completion of event $\sigma$ transmission; state 1, on the other
hand, is unmarked because the transmission is still ongoing.

The capacity of channel ${\bf CH}({j,\sigma,i})$ is 1, meaning that
only when the latest occurrence of event $\sigma$ is received by its
recipient ${\bf G}_i$, will the channel accept a fresh instance of
$\sigma$ from ${\bf G}_j$. Hence, ${\bf CH}({j,\sigma,i})$ permits
reoccurrence of $\sigma$ (i.e. ${\bf G}_j$ sends $\sigma$ again)
only when it is idle, namely at state 0. The capacity constraint of
${\bf CH}({j,\sigma,i})$ can be easily relaxed to allow {\em
multi-capacity} channel models, as we shall see in
Remark~\ref{rem:multicapacity} below. We nevertheless adopt ${\bf
CH}({j,\sigma,i})$ for its structural simplicity and suitability for
clarifying the concept of delay-robustness presented next.


With the channel model ${\bf CH}({j,\sigma,i})$,
we may describe the \emph{channeled behavior} of the system as
follows.  Suppose given ${\bf G}_k$, $k \in \mathcal {N}$; by localization
(see Sect.~\ref{SubSec:SuperviLocali}) ${\bf G}_k$ acquires a set of
local controllers ${\bf LOC}_{k}^C:=\{{\bf LOC}_{\alpha}^C|\alpha \in
\Sigma_{k,hib}\}$ and a set of local preemptors ${\bf
LOC}_{k}^P:=\{{\bf LOC}_{\beta}^P|\beta \in \Sigma_{k,for}\}$.\footnote{\label{foot:selfloop}
For each state state $x$ of each controller ${\bf LOC}_\alpha^C$ (resp. preemptor
${\bf LOC}_\beta^P$), and each communication event $\sigma \in \Sigma_\alpha - \Sigma_k$
(resp. $\sigma \in \Sigma_\beta - \Sigma_k$), if $\sigma$ is not defined at $x$,
we add a $\sigma$-selfloop, i.e. transition $(x, \sigma, x)$ to ${\bf LOC}_\alpha^C$
(resp. ${\bf LOC}_\beta^P$). Now, $\sigma$ is defined at every state of ${\bf LOC}_\alpha^C$
(resp. ${\bf LOC}_\beta^P$). With this modification, the new local controllers
${\bf LOC}_\alpha^C$ (resp. local preemptors ${\bf LOC}_\beta^P$) are also control equivalent
to SUP (because ${\bf LOC}_\alpha^C$ (resp. ${\bf LOC}_\beta^P$) does not disable events $\sigma$ from other components
${\bf G}_j$) and the definition of $\sigma$ at every state of ${\bf LOC}_\alpha^C$
(resp. ${\bf LOC}_\beta^P$) is consistent with the assumption that ${\bf LOC}_\alpha^C$
(resp. ${\bf LOC}_\beta^P$) may receive $\sigma$ after indefinite communication delay.} So
the local controlled behavior of ${\bf G}_k$ is
\begin{align} \label{eq:LocalBehavi}
{\bf SUP}_k := {\bf G}_k
~\mathop{||}~\big(\mathop{||}\limits_{\alpha\in \Sigma_{k,hib}}{\bf
LOC}_\alpha^C\big)
                    ~\mathop{||}~\big(\mathop{||}\limits_{\beta\in \Sigma_{k,for}}{\bf
                    LOC}_\beta^P\big).
\end{align}
Observe that when ${\bf G}_j$ sends $\sigma$ to ${\bf G}_i$ through
${\bf CH}({j,\sigma,i})$, only the recipient ${\bf G}_i$'s local
behavior ${\bf SUP}_i$ is affected because ${\bf G}_i$ receives
$\sigma'$ instead of $\sigma$ due to delay.  Hence each transition
$\sigma$ of ${\bf SUP}_i$ must be replaced by its signal event
$\sigma'$; we denote by ${\bf SUP}'_i$ the resulting new local
behavior of ${\bf G}_i$.  Now let
\begin{align}\label{eq:ModuBehaviNoConnect}
{\bf NSUP} := {\bf SUP}_i'~\mathop{||}~(\mathop{||}\limits_{k\in
\mathcal {N},k\neq i}{\bf SUP}_k)
\end{align}
and then
\begin{align}\label{eq:NewModuBehavi}
{\bf SUP'} := {\bf NSUP}~\mathop{||}~{\bf CH}({j,\sigma,i}).
\end{align}
So ${\bf SUP'}$ is the channeled behavior of the system with respect
to ${\bf CH}({j,\sigma,i})$. Note that both ${\bf SUP'}$ and ${\bf
NSUP}$ are defined over $\Sigma' := \Sigma \cup \{\sigma'\}$.

Let $P:\Sigma'^* \rightarrow \Sigma^*$ and $P_{ch}:\Sigma'^*
\rightarrow \{\sigma,tick,\sigma'\}^*$ be natural projections (as in
(\ref{eq:natpro})). We define delay-robustness as follows.

\begin{defn}\label{defn:unboundDelayRobust}
Consider that ${\bf G}_j$ sends event $\sigma$ to ${\bf G}_i$
through channel ${\bf CH}({j,\sigma,i})$. The monolithic supervisor
${\bf SUP}$ in (\ref{eq:MonoliSUP}) is {\it delay-robust} with
respect to ${\bf CH}({j,\sigma,i})$ if the following conditions
hold:

\noindent (i) ${\bf SUP'}$ in (\ref{eq:NewModuBehavi}) is
\emph{correct} and \emph{complete}, i.e.
\begin{align}
&PL({\bf SUP'}) = L({\bf SUP}) \label{eq:sub1:DefDelayRobust}\\
&PL_m({\bf SUP'}) = L_m({\bf SUP}) \label{eq:sub2:DefDelayRobust}\\
&(\forall s \in \Sigma'^*)(\forall w \in \Sigma^*) ~s \in L({\bf SUP'})~ \& ~(Ps)w \in L_m({\bf SUP})\notag\\
&~~~~~~~~~~~~\Rightarrow (\exists v \in \Sigma'^*)~Pv = w ~\& ~sv
\in L_m({\bf SUP'}) \label{eq:sub3:DefDelayRobust}
\end{align}

\noindent (ii) $P^{-1}_{ch} (L({\bf CH}({j,\sigma,i})))$ is
controllable with respect to $L({\bf NSUP})$ and  $\{\sigma\}$, i.e.
\begin{align} \label{eq:ControllabilityTest}
P_{ch}^{-1}L({\bf CH}({j,\sigma,i}))\{\sigma\} \ \cap \ L({\bf
NSUP}) \subseteq P_{ch}^{-1}L({\bf CH}({j,\sigma,i}))
\end{align}
\end{defn}

In condition (i) above, `correctness' of ${\bf SUP'}$ means that no
$P$-projection of anything ${\bf SUP}'$ can do is disallowed by $\bf
SUP$, while `completeness' means that anything $\bf SUP$ can do is
the $P$-projection of something $\bf SUP'$ can do. In this sense,
the channeled behavior ${\bf SUP'}$ is `equivalent' to its
delay-free counterpart ${\bf SUP}$.  Specifically, conditions
(\ref{eq:sub1:DefDelayRobust}) and (\ref{eq:sub2:DefDelayRobust})
state the equality of closed and marked behaviors between $\bf SUP$
and the $P$-projection of $\bf SUP'$; condition
(\ref{eq:sub3:DefDelayRobust}), which is required for
`completeness', states that if ${\bf SUP}'$ executes a string $s$
whose projection $Ps$ in $\bf SUP$ can be extended by a string $w$
to a marked string of $\bf SUP$, then ${\bf SUP}'$ can further
execute a string $v$ whose projection $Pv$ is $w$ and such that $sv$
is marked in ${\bf SUP}'$.  Roughly, an {\it observationally
consistent inference} about coreachability at the ``operating''
level of ${\bf SUP}'$ can be drawn from coreachability at the
abstract (projected) level of ${\bf SUP}$.


Condition (ii) of Definition~\ref{defn:unboundDelayRobust} imposes a
basic requirement that channel ${\bf CH}({j,\sigma,i})$, when
combined with {\bf NSUP} in (\ref{eq:ModuBehaviNoConnect}) to form
${\bf SUP'}$, should not entail uncontrollability with respect to
$\sigma$. We impose condition~(ii) no matter whether
$\sigma$ is controllable or uncontrollable. This is because we view
the channel ${\bf CH}(j,\sigma,i)$ as a hard-wired passive adjunction
to the original system, and therefore ${\bf CH}(j,\sigma,i)$ cannot
exercise control on $\sigma$. In other words, the channel
has to `accept' any event that the rest of the system might execute, whether
that event is controllable or uncontrollable. Thus if there is already
an instance of $\sigma$ in the channel (i.e. ${\bf CH}(j,\sigma,i)$ at
state 1), then reoccurrence of $\sigma$ will be (unintentionally) `blocked',
causing condition (ii) to fail.
This issue persists, albeit in milder form,
even if we use channel models of multiple (finite) capacities (see
Remark~\ref{rem:multicapacity} below).

We note that delay-robustness as defined above is an extension, from untimed DES
to timed DES, of the concept proposed under the same name in \cite{ZhangCai:conf12} .
In particular, the channel model ${\bf CH}({j,\sigma,i})$ used in the
definition is capable of measuring transmission delay by counting $tick$
occurrences; and condition (ii) in the definition requires controllability for timed DES.

Finally, we present a polynomial algorithm to verify the
delay-robustness property. Notice that when
(\ref{eq:sub1:DefDelayRobust}) and (\ref{eq:sub2:DefDelayRobust})
hold, then (\ref{eq:sub3:DefDelayRobust}) is identical with the {\it
$L_m({\bf SUP}')$-observer} property of $P$
\cite{WongWonham:2004,FenWon:08}. The latter may be verified in
polynomial time ($O(n^4)$, $n$ the state size of ${\bf SUP}'$) by
computing the \emph{supremal quasi-congruence} of a nondeterministic
automaton derived from ${\bf SUP}'$ and $P$
\cite{WongWonham:2004,FengWonham:2010}.\footnote{We note \emph{en passant} that
\cite{Bravo:2012} reports an algorithm with quadratic time complexity for verifying
the observer property alone; that does not, however, yield
structural information which (if the observer property is not satisfied) might be
useful for remedial design.} The following is the
delay-robustness verification algorithm.

\noindent \textbf{Algorithm~1}

\noindent 1. Check if $P$ is an $L_m({\bf SUP}')$-observer. If no,
return \emph{false}.

\noindent 2. Check if $PL({\bf SUP'}) = L({\bf SUP})$ and $PL_m({\bf
SUP'}) = L_m({\bf SUP})$. If no, return \emph{false}.

\noindent 3. Check if $P^{-1}_{ch} (L({\bf CH}({j,\sigma,i})))$ is
controllable with respect to $L({\bf NSUP})$ and $\{\sigma\}$.  If
no, return \emph{false}.

\noindent 4. Return \emph{true}.

If Step~1 above ($O(n^4)$ complexity) is successful, i.e. $P$ is
indeed an $L_m({\bf SUP}')$-observer, then Step~2 of computing
$PL({\bf SUP'})$ and $PL_m({\bf SUP'})$ is of polynomial complexity
$O(n^4)$ \cite{FengWonham:2010}. Then checking the two equalities in
Step~2 is of $O(n^2)$ complexity. Finally in Step~3, controllability
may be checked using standard algorithm \cite{BrandinWonham:94} in
linear time $O(n)$. Therefore, Algorithm~1 terminates and is of
polynomial complexity $O(n^4)$. The following result is
straightforward.

\begin{pro}\label{pro:VerifUDR}
Consider that ${\bf G}_j$ sends event $\sigma$ to ${\bf G}_i$
through channel ${\bf CH}({j,\sigma,i})$. The monolithic supervisor
${\bf SUP}$ is {\it delay-robust} with respect to ${\bf
CH}({j,\sigma,i})$ if and only if Algorithm~1 returns true.
\end{pro}

\begin{figure}[!t]
\centering
    \includegraphics[scale = 0.6]{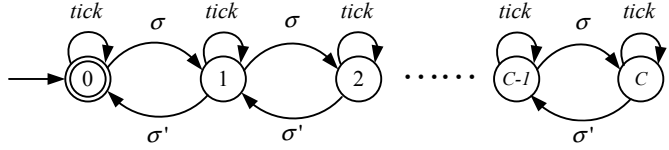}
\caption{$C$-capacity channel model ${\bf NCH}({j,\sigma,i})$.}
\label{fig:Ch_C}
\end{figure}

\begin{remark} \label{rem:multicapacity} (Multi-capacity channel model)
So far we have considered the 1-capacity channel model ${\bf
CH}({j,\sigma,i})$, and defined delay-robustness with respect to it.
We now consider the more general $C$-capacity channel model ${\bf
NCH}({j,\sigma,i})$, $C \geq 1$ a positive integer, displayed in
Fig.~\ref{fig:Ch_C}. The sender ${\bf G}_j$ may send at most $C$
instances of event $\sigma$ to ${\bf NCH}({j,\sigma,i})$, each
instance subject to indefinite delay. With channel ${\bf
NCH}({j,\sigma,i})$, one may proceed just as before, by
replacing ${\bf CH}({j,\sigma,i})$ by ${\bf NCH}({j,\sigma,i})$
throughout, to define the corresponding delay-robustness property
with respect to ${\bf NCH}({j,\sigma,i})$, and then revising
Algorithm~1 correspondingly to verify delay-robustness.

It is worth noting that when ${\bf NCH}({j,\sigma,i})$ reaches its
maximal capacity, and ${\bf G}_j$ sends yet another instance of
$\sigma$, then $\sigma$ is `blocked' by ${\bf NCH}({j,\sigma,i})$,
implying uncontrollability of the channeled behavior. Hence the
uncontrollability problem always exists as long as the channel model
is of finite capacity and delay is indefinite, although the
controllability condition (cf. condition (ii) of
Definition~\ref{defn:unboundDelayRobust}) is more easily satisfied
for larger capacity channels (simply because more instances of
$\sigma$ may be sent to the channel).
\end{remark}

\section{Bounded Delay-Robustness and Maximal Delay Bound}\label{Sec:MaxDelayBound}



Consider again the situation that agent ${\bf G}_j$ sends an event
$\sigma$ to ${\bf G}_i$.  If the monolithic supervisor ${\bf SUP}$
is verified (by Algorithm~1) to be delay-robust, then we will use
channel ${\bf CH}({j,\sigma,i})$ in Fig.~\ref{fig:ComChannel} to
transmit $\sigma$ subject to unbounded delay, and the system's behavior
will not be affected. If, however, ${\bf SUP}$ fails to be
delay-robust, there are two possible implications: (1) $\sigma$ must
be transmitted without delay (as in the original setup of
localization \cite{CaiWonham:2010a,ZhangCai:2013,CaiZhangWonham:2013}); or (2) there
exists a delay bound $d$ ($\geq 1$) of $\sigma$ such that if each
transmission of $\sigma$ is completed within $d$ occurrences of
$tick$, the system's behavior will remain unaffected.  This section aims
to identify the latter case, which we call ``bounded delay-robust'',
and moreover to determine the bound $d$.

\begin{figure}[!t]
\centering
    \includegraphics[scale = 0.6]{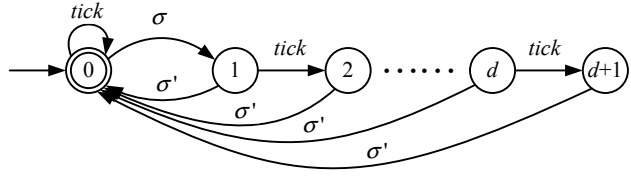}
\caption{Timed channel model ${\bf CH}_d({j,\sigma,i})$, $d \geq 1$,
for transmitting event $\sigma$ from ${\bf G}_j$ to ${\bf G}_i$ with
delay bound $d$.} \label{fig:BoundComChannel}
\end{figure}

To that end, consider the channel model ${\bf CH}_d({j,\sigma,i})$
in Fig.~\ref{fig:BoundComChannel}, with parameter $d \geq 1$. ${\bf
CH}_d({j,\sigma,i})$ is a $(d+2)$-state TDES with event set
$\{\sigma, tick, \sigma'\}$.  After an occurrence of $\sigma$ (state
0 to 1), ${\bf CH}_d({j,\sigma,i})$ counts up to $d$ ($\geq 0$) occurrences of $tick$
(state 1 through $d+1$) by which time the signal event $\sigma'$ must
occur. That is, the occurrence of $\sigma'$ (${\bf G}_i$ receives $\sigma$)
is bounded by $d$ $tick$s. Note that the $tick$ selfloop at state 0
is again for the technical requirement to prevent the blocking of event $tick$
when synchronizing ${\bf CH}_d(j,\sigma,i)$ with other TDES.

Now with ${\bf CH}_d({j,\sigma,i})$, the channeled behavior of the
system is
\begin{align}\label{eq:BNewModuBehavi}
{\bf SUP'}_d := {\bf NSUP}~\mathop{||}~{\bf CH}_d({j,\sigma,i})
\end{align}
where ${\bf NSUP}$ is given in (\ref{eq:ModuBehaviNoConnect}). The
event set of ${\bf SUP'}_d$ is $\Sigma' = \Sigma \cup \{\sigma'\}$,
and we recall the natural projections $P:\Sigma'^* \rightarrow
\Sigma^*$ and $P_{ch}:\Sigma'^* \rightarrow
\{\sigma,tick,\sigma'\}^*$.

\begin{defn}\label{defn:BoundDR}
Consider that ${\bf G}_j$ sends event $\sigma$ to ${\bf G}_i$
through channel ${\bf CH}_d({j,\sigma,i})$, $d \geq 1$. The
monolithic supervisor ${\bf SUP}$ in (\ref{eq:MonoliSUP}) is {\it
bounded delay-robust} with respect to ${\bf CH}_d({j,\sigma,i})$ (or
{\it d-bounded delay-robust}) if the following conditions hold:

\noindent (i) ${\bf SUP'}_d$ in (\ref{eq:BNewModuBehavi}) is
\emph{correct} and \emph{complete}, i.e.
\begin{align}
&PL({\bf SUP'}_d) = L({\bf SUP}) \label{eq:sub1:DefBDelayRobust}\\
&PL_m({\bf SUP'}_d) = L_m({\bf SUP}) \label{eq:sub2:DefBDelayRobust}\\
&(\forall s \in \Sigma'^*)(\forall w \in \Sigma^*) ~s \in L({\bf SUP'}_d)~ \& ~(Ps)w \in L_m({\bf SUP})\notag\\
&~~~~~~~~~~~~\Rightarrow (\exists v \in \Sigma'^*)~Pv = w ~\& ~sv
\in L_m({\bf SUP'}_d) \label{eq:sub3:DefBDelayRobust}
\end{align}

\noindent (ii) $P^{-1}_{ch} (L({\bf CH}_d({j,\sigma,i})))$ is
controllable with respect to $L({\bf NSUP})$ and $\{\sigma\}$, i.e.
\begin{align} \label{eq:BControllabilityTest}
P_{ch}^{-1}L({\bf CH}_d({j,\sigma,i}))\{\sigma\} \ \cap \ L({\bf
NSUP}) \subseteq P_{ch}^{-1}L({\bf CH}_d({j,\sigma,i}))
\end{align}
\end{defn}

Bounded delay-robustness is defined in the same way as (unbounded)
delay-robustness in Definition~\ref{defn:unboundDelayRobust}, but
with respect to the new channel model ${\bf CH}_d({j,\sigma,i})$
with delay bound $d$. As a result, $d$-bounded delay-robustness may
be verified by Algorithm~1 with corresponding modifications. For
later reference, we state here the modified algorithm.

\noindent \textbf{Algorithm~2}

\noindent 1. Check if $P$ is an $L_m({\bf SUP}'_d)$-observer. If not,
return \emph{false}.

\noindent 2. Check if $PL({\bf SUP'}_d) = L({\bf SUP})$ and
$PL_m({\bf SUP'}_d) = L_m({\bf SUP})$. If not, return \emph{false}.

\noindent 3. Check if $P^{-1}_{ch} (L({\bf CH}_d({j,\sigma,i})))$ is
controllable with respect to $L({\bf NSUP})$ and $\{\sigma\}$.  If
not, return \emph{false}.

\noindent 4. Return \emph{true}.


Now if the monolithic supervisor ${\bf SUP}$ fails to be (unbounded)
delay-robust with respect to channel ${\bf CH}({j,\sigma,i})$, we
would like to verify if ${\bf SUP}$ is bounded delay-robust with
respect to ${\bf CH}_d({j,\sigma,i})$ for some $d \geq 1$.  If so,
compute the \emph{maximal} delay bound, i.e. the largest delay (number
of $tick$s) that can be tolerated without changing the system's logical behavior. We
need the following lemma.

\begin{lem} \label{pro:BoundRelat}
Consider that ${\bf G}_j$ sends event $\sigma$ to ${\bf G}_i$
through channel ${\bf CH}_d({j,\sigma,i})$, $d \geq 1$. If ${\bf
SUP}$ is not $d$-bounded delay-robust, then it is not
$(d+1)$-bounded delay-robust.
\end{lem}

The result of Lemma~\ref{pro:BoundRelat} is intuitive: if ${\bf
SUP}$ cannot tolerate a $\sigma$ transmission delay of $d$, neither
can it tolerate a delay $(d+1)$. By induction, in fact, ${\bf SUP}$
cannot tolerate any delay larger than $d$.  The proof of
Lemma~\ref{pro:BoundRelat} is in Appendix~\ref{app:BoundDR}. This
fact suggests the following algorithm for identifying bounded
delay-robustness as well as computing the maximal delay bound.

\noindent \textbf{Algorithm~3}

\noindent 1. Set $d = 1$.

\noindent 2. Check by Algorithm~2 if $\bf SUP$ is $d$-bounded
delay-robust relative to channel ${\bf CH}_d({j,\sigma,i})$. If not,
let $d = d - 1$ and go to Step 3. Otherwise advance $d$ to $d + 1$
and repeat Step 2.

\noindent 3. Output $d_{max} := d$.

\begin{lem} \label{pro:Alg3FiniteTerminate}
If $\bf SUP$ is not delay-robust with respect to ${\bf CH}(j,\sigma,i)$, then Algorithm~3 terminates
in at most $2^m*m$ steps, i.e. $d_{max} \leq 2^m*m$, where m is the state size of ${\bf SUP}'$ in (\ref{eq:NewModuBehavi}).
\end{lem}

The proof of Lemma~\ref{pro:Alg3FiniteTerminate} is given in
Appendix B. In Algorithm~3, we work upwards starting from the minimal
delay $d=1$. If $\bf SUP$ is {\em not} $1$-bounded delay-robust with
respect to ${\bf CH}_1({j,\sigma,i})$, then by
Lemma~\ref{pro:BoundRelat} $\bf SUP$ is {\em not} $d$-bounded
delay-robust for any $d > 1$. Therefore $\bf SUP$ is {\em not}
bounded delay-robust and $\sigma$ must be transmitted without delay.
Note that in this case Algorithm~3 outputs $d_{max}=0$.

If $\bf SUP$ is $1$-bounded delay-robust, we next check if it is
$2$-bounded delay-robust with respect to ${\bf CH}_2({j,\sigma,i})$.
If $\bf SUP$ fails to be $2$-bounded delay-robust, then again by
Lemma~\ref{pro:BoundRelat} $\bf SUP$ fails to be $d$-bounded
delay-robust for any $d > 2$. Hence $\bf SUP$ is bounded
delay-robust, with the maximal delay bound $d_{max}=1$.

If $\bf SUP$ is shown to be $2$-bounded delay-robust, the iterative
process continues until $\bf SUP$ fails to be $(d+1)$-bounded
delay-robust for some $d \geq 2$; this happens in finitely many
steps according to Lemma~\ref{pro:Alg3FiniteTerminate}. Then $\bf
SUP$ is bounded delay-robust, with the maximal delay bound
$d_{max}=d$. The following result is immediate.

\begin{pro}\label{pro:VerifBDR}
Consider that ${\bf G}_j$ sends event $\sigma$ to ${\bf G}_i$
through channel ${\bf CH}_d({j,\sigma,i})$, $d \geq 1$. The
monolithic supervisor ${\bf SUP}$ is bounded delay-robust with
respect to ${\bf CH}_d({j,\sigma,i})$ if and only if the output
$d_{max}$ of Algorithm~3 satisfies $d_{max} > 0$.  Moreover, if
${\bf SUP}$ is bounded delay-robust, then $d_{max}$ is the maximal
delay bound for $\sigma$ transmission.
\end{pro}

To summarize, when an event $\sigma$ is sent from ${\bf G}_j$ to
${\bf G}_i$, we determine unbounded or bounded delay-robustness and
choose the corresponding channel as follows.

\noindent \textbf{Algorithm~4}

\noindent 1. Check by Algorithm~1 if $\bf SUP$ is (unbounded)
delay-robust. If so, terminate, set the maximal delay bound $d_{max}
= \infty$, and use channel ${\bf CH}({j,\sigma,i})$ in
Fig.~\ref{fig:ComChannel}.

\noindent 2. Check by Algorithm~3 if $\bf SUP$ is bounded
delay-robust. If so (i.e. $d_{max} \geq 1$), terminate and use
channel ${\bf CH}_d({j,\sigma,i})$ in Fig.~\ref{fig:BoundComChannel}
with $d = d_{max}$.

\noindent 3. In this case $d_{max}=0$. Terminate and use no channel:
$\sigma$ must be transmitted without delay.

\begin{remark} \label{rem:multieventcomm} (Multiple channeled events)
So far we have considered a single event communication: agent ${\bf
G}_j$ sends event $\sigma$ to ${\bf G}_i$. Using this as a basis, we
present an approach to the general case of multiple
channeled events, as is common in distributed control.
We will consider that each fixed triple (sender,
channeled event, receiver) is assigned with its own communication
channel, and the assigned channels operate concurrently. Our goal is
to obtain these channels, ensuring unbounded or bounded
delay-robustness, one for each triple (sender, channeled event,
receiver).

First fix $i,j \in \mathcal {N}$, and recall from
(\ref{eq:ChnEvent}) that $\Sigma_{j,com,i}$ is the set of events
that ${\bf G}_j$ sends to ${\bf G}_i$.  Write $\Sigma_{j,com,i} =
\{\sigma_1,...,\sigma_r\}$, $r \geq 1$, and treat the channeled events $\sigma_1$,
$\sigma_2$, ... {\it sequentially}, in order of indexing.

\noindent \textbf{Algorithm~5}

\noindent 1. Set $p = 1$.

\noindent 2. For event $\sigma_p \in \Sigma_{j,com,i}$ apply
Algorithm~4 to obtain the maximal delay bound $d_{max}$.

2.1. If $d_{max}=\infty$, namely unbounded delay-robustness, choose
channel ${\bf CH}({j,\sigma_p,i})$, and let ${\bf NSUP} := {\bf
NSUP} || {\bf CH}({j,\sigma_p,i})$.

2.2. If $d_{max} \geq 1$ is finite, namely bounded delay-robustness,
choose channel ${\bf CH}_d({j,\sigma_p,i})$, and let ${\bf NSUP} :=
{\bf NSUP} || {\bf CH}_d({j,\sigma_p,i})$.

2.3 If $d_{max} = 0$, then no channel is chosen and $\sigma_p$ must
be transmitted without delay.

\noindent If $p < r$, advance $p$ to $p+1$ and repeat Step~2.

\noindent 3. Output a set of channels used for sending events from
${\bf G}_j$ to ${\bf G}_i$.

Note that at Step~2 of Algorithm~5, if a channel is chosen for event
$\sigma_p$, then ${\bf NSUP}$ must be reset to be the synchronous
product of ${\bf NSUP}$ and the channel, so that in choosing a
channel for the next event $\sigma_{p+1}$ the previously chosen
channel is considered together. This ensures that when the derived
channels operate concurrently, the system's behavior is not
affected. It is worth noting that a different ordering of the set
$\Sigma_{j,com,i}$ may result in a different set of channels; if no priority of
the transmission delay is imposed on the communication events, we may choose an
ordering randomly.

Finally, since the set of all communication events is $\Sigma_{com}
:= \mathop{\cup}\limits_{i,j\in \mathcal {N}}\Sigma_{j,com,i}$, we
simply apply Algorithm~5 for each (ordered) pair $i, j \in \mathcal
{N}$ to derive all communication channels. Again, a different ordering of the
set $\mathcal{N}\times \mathcal{N}$ generally results in a different
set of channels, because the channels chosen for a pair $(i,j)$ will
be used to decide channels for all subsequent $(i', j')$. For convenience
we will simply order the pairs $(i,j)$ sequentially first on $j$ then on $i$.
\end{remark}


\section{Case Study: Under-Load Tap-Changing Transformer} \label{Sec:DSCofULTC}
In this section we demonstrate timed delay-robustness and associated verification algorithms on an under-load tap-changing transformer system.

\subsection{Model Description and Supervisor Localization} \label{ULTCDistributedControl}

\begin{figure}[!t]
\begin{center}
\includegraphics[scale = 0.8]{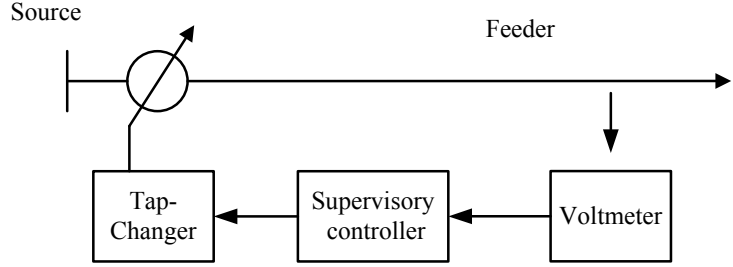}
\caption{ULTC: Components and Controller}
\label{fig:ULTC}
\end{center}
\end{figure}

Transformers with tap-changing facilities constitute an important means of controlling voltage at all levels throughout electrical power systems. We consider an under-load tap-changing transformer (ULTC) as displayed in Fig.~\ref{fig:ULTC}, which consists of two components: Voltmeter and Tap-Changer\cite{AfzaSaadWonham:2008}.

This ULTC is operated in two modes: Automatic and Manual.
In the automatic mode, the tap-changer works according to the following logic. (1) If the voltage deviation is greater than some threshold value, then a timer will start; when the timer times out, a `tap increase (or decrease) event' will occur and the timer will reset; a tap increase or decrease
should only occur if the voltage change continues to exceed threshold after the time out- this is to avoid tap changes in response to merely occasional random fluctuations of brief duration.
(2) If the voltage returns to the dead-band, because of a tap change or some other reason, then no tap change will occur.
(3) If the voltage exceeds the maximally allowed value $V_{max}$, then lowering of the tap command without delay occurs instantaneously.
In the manual mode, the system is waiting for `Tap-up', `Tap-down', or `Automatic' commands. An operator can change the operation mode from one to the other, and thus the operator is adjoined into the plant components to be controlled.

\begin{figure}[!t]
\begin{center}
\includegraphics[scale = 0.6]{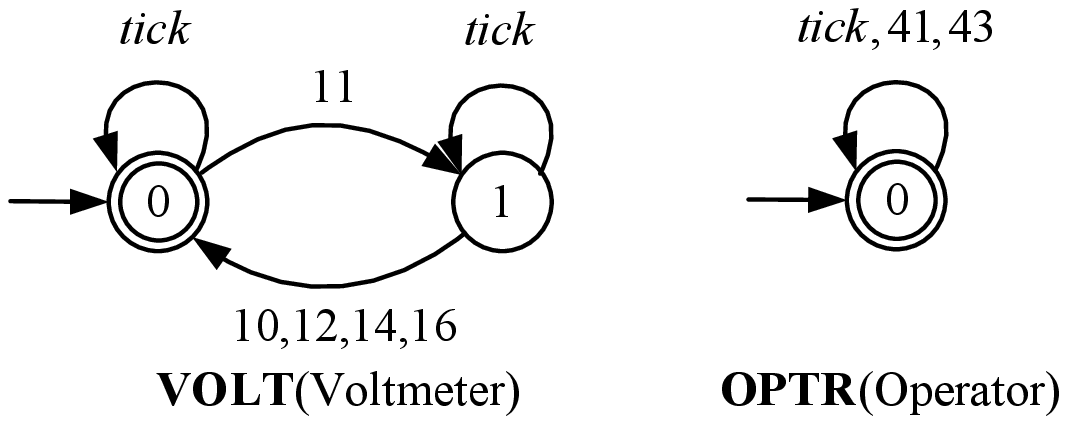}\\
\vspace{2em}
\includegraphics[scale = 0.7]{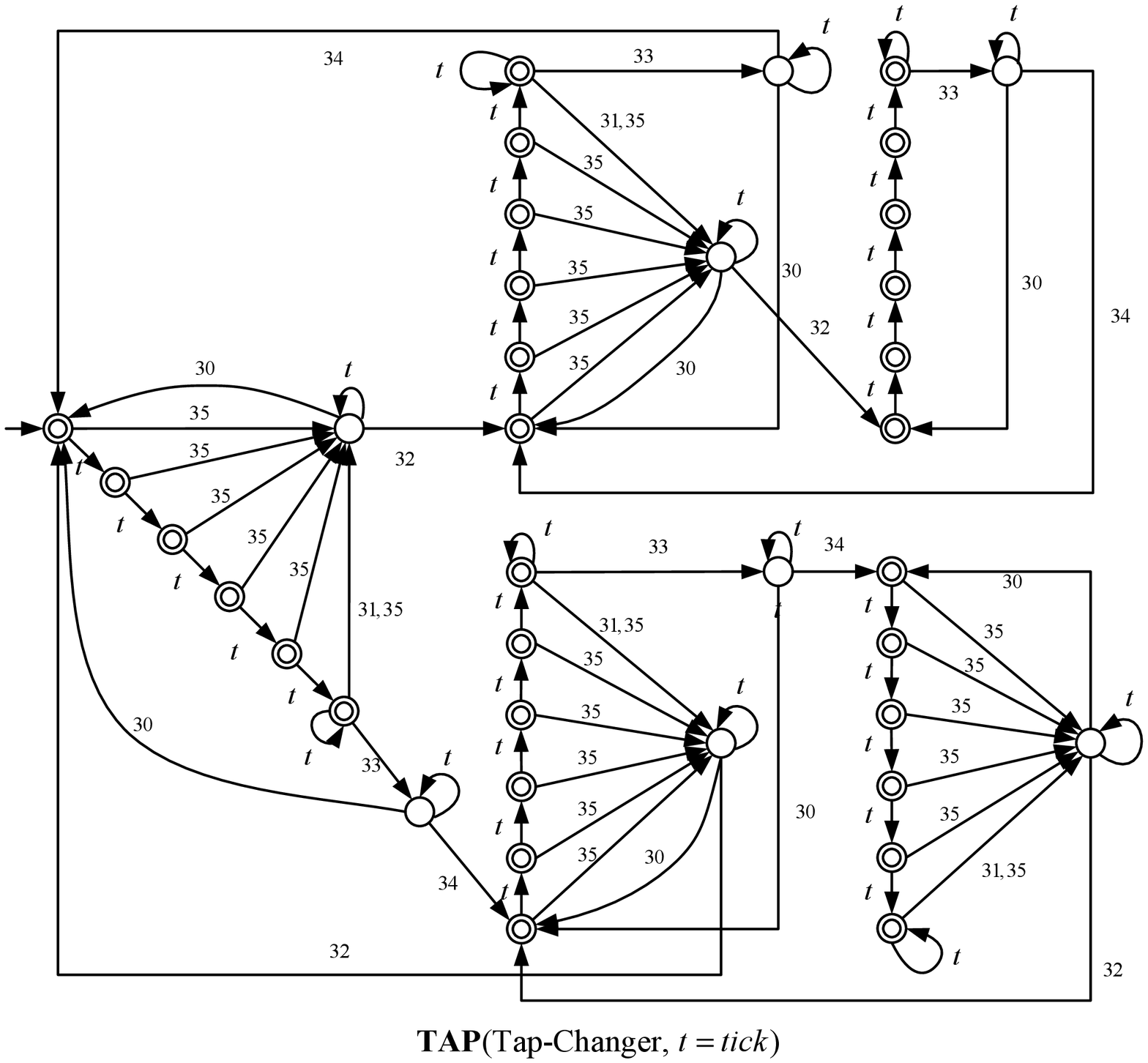}
\caption{Timed Transition Graph of ULTC Components}
\label{fig:TDES Model}
\end{center}
\end{figure}

Each plant component is modeled as a TDES displayed in Fig.~\ref{fig:TDES Model}, and associated events are listed in Table~\ref{tab:event}. So, the plant to be controlled is the synchronized behavior 
of Voltmeter ($\bf VOLT$), Tap-changer ($\bf TAP$) and Operator ($\bf OPTR$), i.e.
\begin{align} \label{eq:ULTC:Plant}
{\bf PLANT} = {\bf VOLT}~||~ {\bf TAP} ~||~ {\bf OPTR}.
\end{align}

\begin{table}
\caption{Physical interpretation of events}
\label{tab:event}
\begin{center}
\scalebox{0.86}{
\begin{tabular}{|c||l|c|l|}
\hline
Event& Physical interpretation & Time bounds& (hib/for)\\
                            &  & (lower, upper) &\\
\hline
11 & Initialize voltmeter & (0, $\infty$) & hib \\
\hline
10 & Report $|\Delta V| > ID$ and $\Delta V > 0$ & (0, $\infty$)&\\
\hline
12 & Report $|\Delta V| < ID$, i.e. voltage recovered & (0, $\infty$) &\\
\hline
14 & Report $|\Delta V| > ID$ and $\Delta V < 0$ & (0, $\infty$) &\\
\hline
16 & Report voltage exceeds $V_{max}$ & (0, $\infty$) & \\
\hline
30 & Tap-up/Down failed & (0, $\infty$) & \\
\hline
31 & Tap-down command with 5 $tick$ delay & (5, $\infty$) & hib \& for\\
\hline
32 & Tap-down successful & (0, $\infty$) & \\
\hline
33 & Tap-up command & (0, $\infty$) & hib \& for \\
\hline
34 & Tap-up successful & (0, $\infty$) &\\
\hline
35 & Tap-down command without delay & (0, $\infty$) & hib \& for\\
\hline
41 & Enter Automatic mode & (0, $\infty$) & hib \\
\hline
43 & Enter Manual mode & (0, $\infty$) & hib\\
\hline
\end{tabular}
}
\end{center}
\end{table}

\begin{figure}[!t]
\begin{center}
\includegraphics[scale = 1.0]{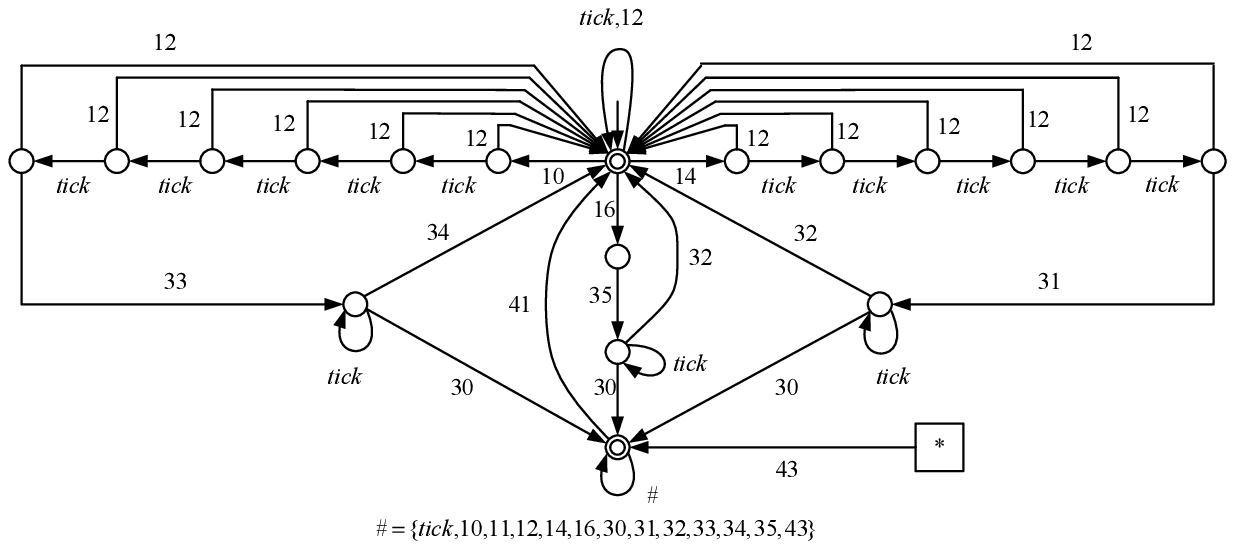}
\caption{Control Specification $\bf SPEC$  in Automatic/Manual Mode. The transition 43
from the square with `*' represents
similar transitions from all states to the `manual operation mode'.}
\label{fig:Specification}
\end{center}
\end{figure}

We consider a voltage control problem of the ULTC: when the voltage is not `normal', design controllers to recover the voltage through controlling tap ratio after a time delay to recover the voltage. Fig.~\ref{fig:Specification} displays the TDES model $\bf SPEC$ for the control specification in Automatic/Manual mode.

Note that since the tap increase (decrease) and lowering tap commands would
preempt the occurrence of $tick$, the corresponding events 31, 33 and 35 are designated as forcible
events. In the following, we synthesize the monolithic supervisor $\bf SUP$ by the standard TDES supervisory control theory \cite{BrandinWonham:94, Wonham:2013a} and the local controllers by TDES supervisor localization \cite{ZhangCai:2013,CaiZhangWonham:2013}.

First, synthesize the monolithic supervisor TDES ${\bf SUP}$ in the usual sense that its marked behavior
\begin{align}\label{eq:ULTC:SUP}
L_m({\bf SUP}) = Sup\mathcal{C}(L_m({\bf SPEC})\cap L_m({\bf VOLT}))
\end{align}
and its closed behavior $L({\bf SUP}) = \overline{L_m({\bf SUP})}$. $\bf SUP$ has 231 states and 543 transitions, and embodies disabling actions for all the prohibitible events and preempting actions relative to ${tick}$ for all the forcible events.

\begin{figure}[!t]
\begin{center}
\includegraphics[scale = 0.6]{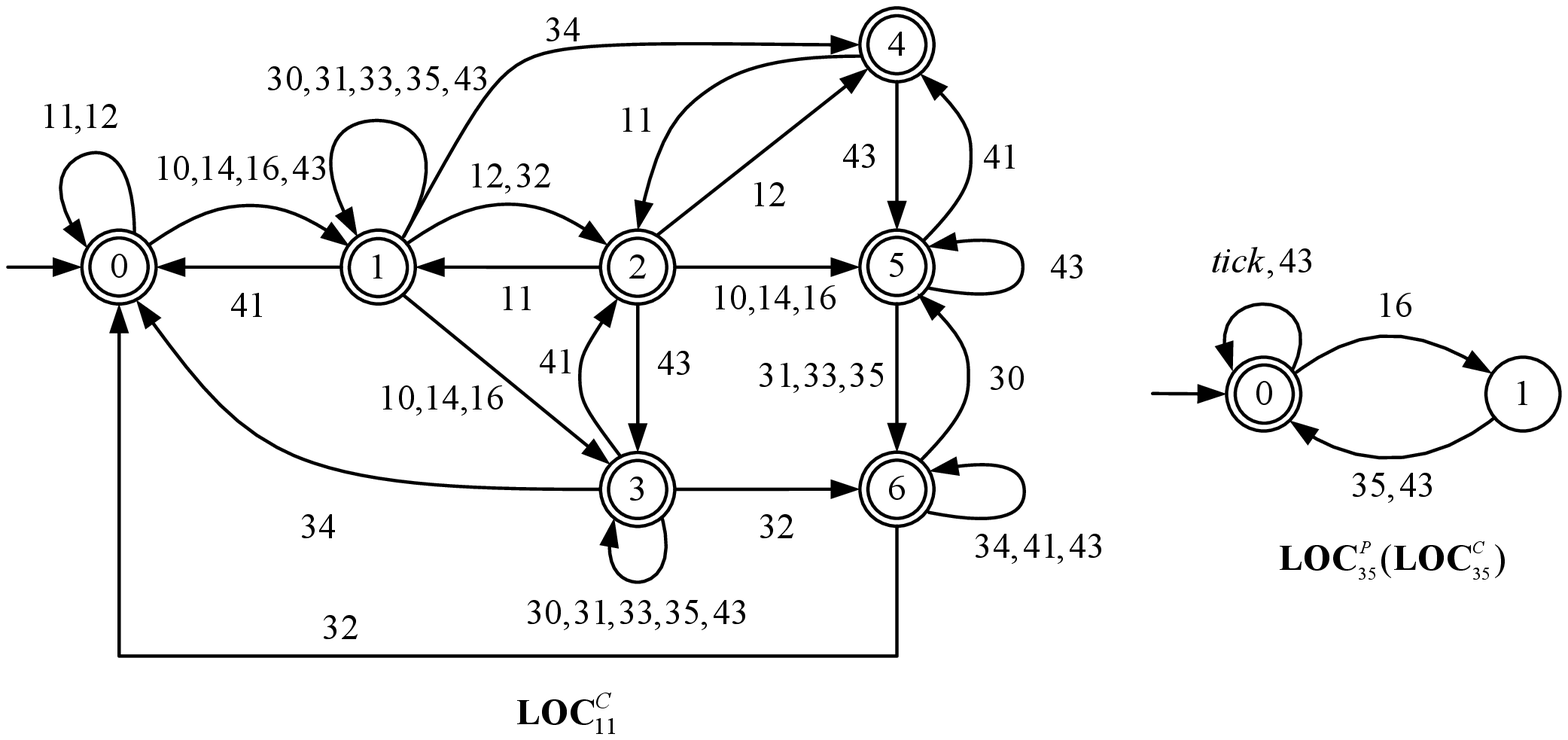}\\
\vspace{0.5em}
\includegraphics[scale = 0.6]{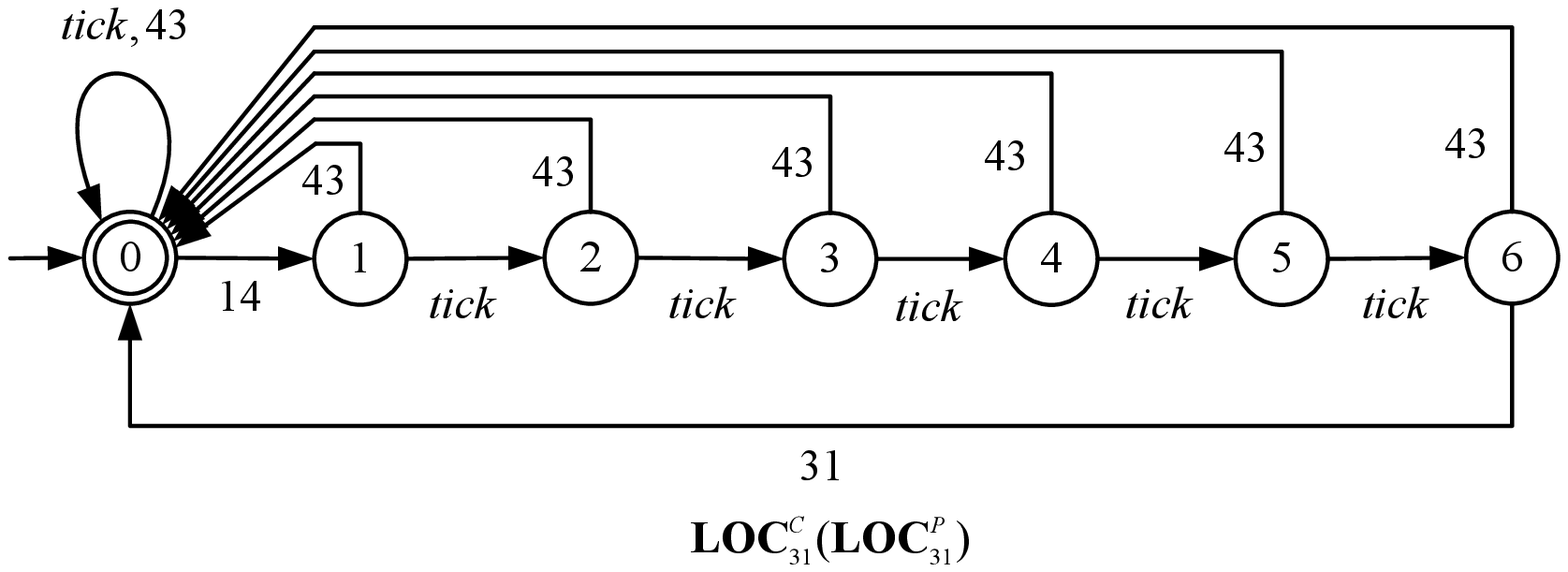}
\hspace{1em}
\includegraphics[scale = 0.6]{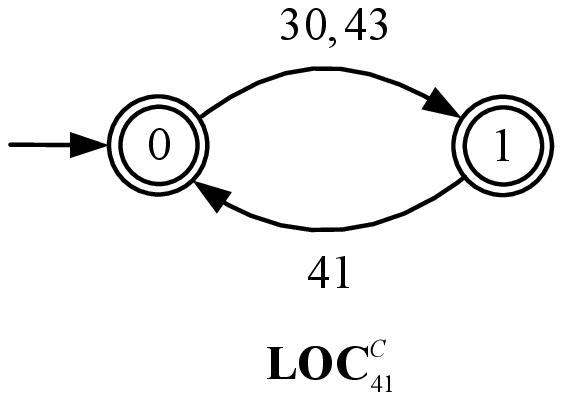}\\
\vspace{0.5em}
\includegraphics[scale = 0.6]{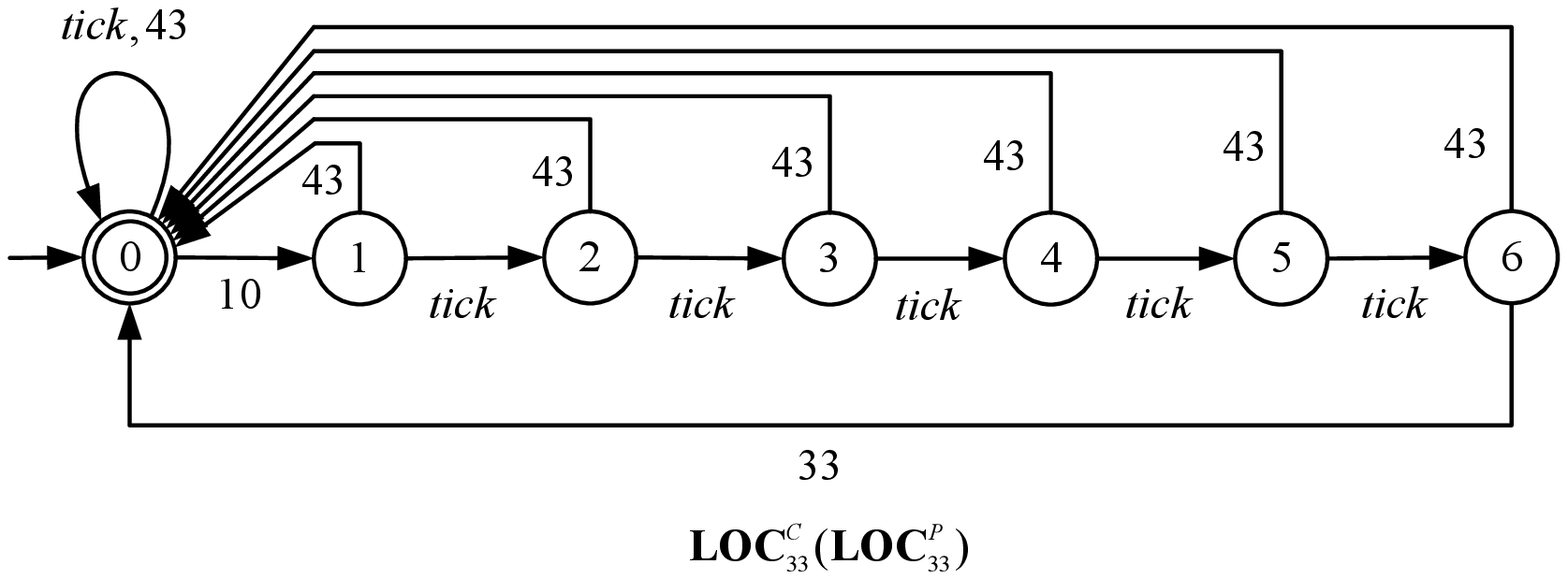}
\hspace{1em}
\includegraphics[scale = 0.6]{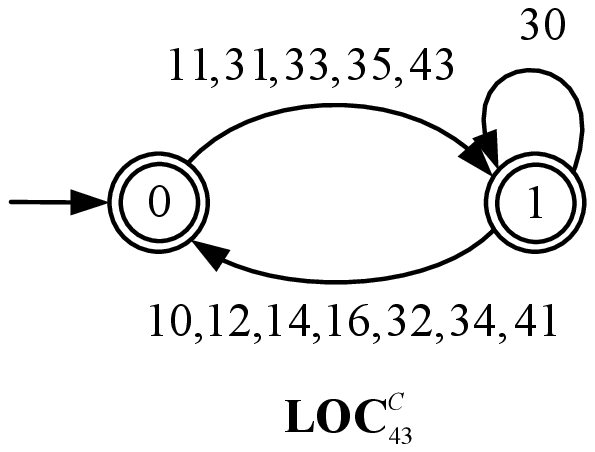}\\
\caption{Local controllers and local preemptors for ULTC. According to Footnote~\ref{foot:selfloop},
for each state state $x$ of each local controller/preemptor, and each communication event $\sigma$,
if $\sigma$ is not defined at $x$, we add a $\sigma$-selfloop. Let *(x) be the set of events whose selfloops
need to be adjoined at state $x$. In ${\bf LOC}_{11}^C$, $*(0) = *(2) = *(4) = \{30,31,32,33,34,35,41\}$, $*(1) = \{43\}$,
$*(5) = \{30,32,34\}$, and $*(6) = \{31,33,35\}$; in ${\bf LOC}_{31}^C$ (${\bf LOC}_{31}^P$), $*(1) = *(2)
= *(3) = *(4) = *(5) = *(6) = \{14\}$; in ${\bf LOC}_{33}^C$ (${\bf LOC}_{33}^P$), $*(1) = *(2)
= *(3) = *(4) = *(5) = *(6) = \{10\}$; in ${\bf LOC}_{35}^C$ (${\bf LOC}_{35}^P$), $*(1) = \{16\}$;
in ${\bf LOC}_{41}^C$, $*(1) = \{30\}$; in ${\bf LOC}_{43}^C$, $*(0) = \{10,12,14,16,30,32,34\}$,
and $*(1) = \{11,31,33,35\}$.}
\label{fig:Local Controllers}
\end{center}
\end{figure}

Next, by supervisor localization, we obtain a set of local controllers ${\bf LOC}_{11}^C$, ${\bf LOC}_{31}^C$, ${\bf LOC}_{33}^C$, ${\bf LOC}_{35}^C$ ${\bf LOC}_{41}^C$ and ${\bf LOC}_{43}^C$ for controllable events 11, 31, 33, 35, 41 and 43 respectively, and a set of local preemptors ${\bf LOC}_{31}^P$, ${\bf LOC}_{33}^P$ and ${\bf LOC}_{35}^P$ for forcible events 31, 33 and 35 respectively; their transition diagrams are shown in Fig.~\ref{fig:Local Controllers}.

\begin{figure}[!t]
\begin{center}
\includegraphics[scale = 0.6]{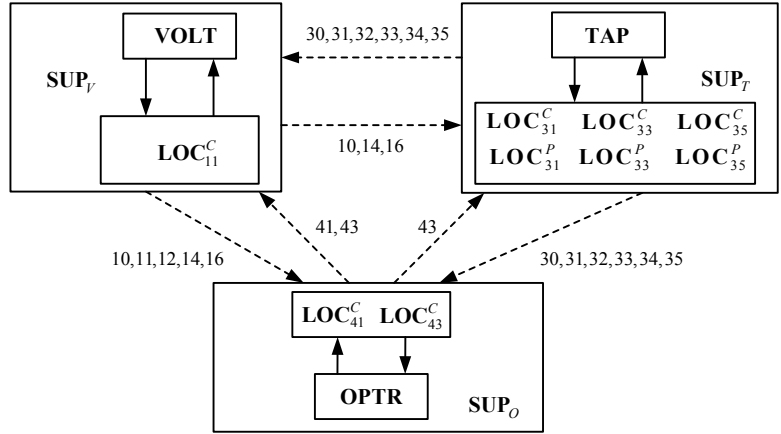}\\
\caption{Distributed Control Architecture of ULTC}
\label{fig:ULTC Distribution}
\end{center}
\end{figure}

Finally, using these constructed local controllers/preemptors, we build a distributed control architecture without communication delays for ULTC as displayed in Fig.~\ref{fig:ULTC Distribution}. The local controlled behaviors of the plant components are
\begin{align*}
{\bf SUP}_V ~=~ &{\bf VOLT}~\mathop{||}~{\bf LOC}_{11}^C,\\
{\bf SUP}_T ~=~ &{\bf TAP}~\mathop{||}~({\bf LOC}_{31}^C ~\mathop{||}~{\bf LOC}_{33}^C~\mathop{||}~{\bf LOC}_{35}^C)\\
&~~~~~~~~\mathop{||}~({\bf LOC}_{31}^P~\mathop{||}~{\bf LOC}_{33}^P~\mathop{||}~{\bf LOC}_{35}^P),\\
{\bf SUP}_O ~=~ &{\bf OPTR}~\mathop{||}~({\bf LOC}_{41}^C~\mathop{||}~{\bf LOC}_{43}^C).
\end{align*}
Let $\Sigma_{A,com,B}$ represent the set of events that component $A$ sends to component $B$; the sets of communication events are
\begin{align}\label{eq:ULTC:ComEvent}
\Sigma_{T,com,V} &= \{30,31,32,33,34,35\}, \notag\\
\Sigma_{O,com,V} &= \{41,43\}, \notag \\
\Sigma_{V,com,T} &= \{10,14,16\}, \notag\\
\Sigma_{O,com,T} &= \{43\}, \\
\Sigma_{V,com,O} &= \{10,11,12,14,16\},\notag\\
\Sigma_{T,com,O} &= \{30,31,32,33,34,35\}. \notag
\end{align}
It is guaranteed by supervisor localization of TDES \cite{ZhangCai:2013,CaiZhangWonham:2013} that the ULTC under the control of these local controllers and preemptors without communication delay, has closed and marked behavior identical to $\bf SUP$ in (\ref{eq:ULTC:SUP}). 

\subsection{Delay-Robustness Verification} \label{SubSec:DRforULTC}


Now we investigate the timed delay-robustness property for ULTC. For illustration, we consider the following three cases.

(1) Event 30 in $\Sigma_{T,com,O}$

\begin{figure}[!t]
\begin{center}
\includegraphics[scale = 0.5]{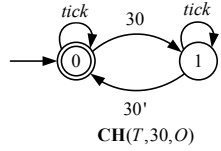}\\
\caption{Communication Channel ${\bf CH}({T,30,O})$}
\label{fig:COM30}
\end{center}
\end{figure}

Applying Algorithm 4, at Step 1 we verify by Algorithm 1 that $\bf SUP$ is delay-robust with respect to the communication channel ${\bf CH}({T,30,O})$ transmitting event 30, as displayed in Fig.~\ref{fig:COM30}.

To illustrate that the overall system behavior will not be affected by indefinite communication delay of event 30, consider the case that the voltmeter reported an increase in voltage (in $\bf VOLT$ as displayed in Fig.~\ref{fig:TDES Model}, events 11 and 10 have occurred), and the tap has received a tap-up command, but the tap-up operation failed (in $\bf TAP$ as displayed in Fig.~\ref{fig:TDES Model}, events ${tick}$, ${tick}$, ${tick}$, ${tick}$, ${tick}$, 33 and 30 have occurred in sequence). By inspection of the transition diagrams, the plant components  $\bf VOLT$, $\bf TAP$ and $\bf OPTR$ in Fig.~\ref{fig:TDES Model} are at states 0, 0, and 0 respectively, and thus the events that are eligible to occur are 11, 35, 41, 43, and $tick$. However, according to the transition diagrams of the local controllers and preemptors displayed in Fig.~\ref{fig:Local Controllers}: (1) ${\bf LOC}_{11}^C$ is at state 1 and disables event 11; (2) ${\bf LOC}_{35}^C$ is at state 0 and disables event 35; (3) ${\bf LOC}_{41}^C$ will disable or enable event 41 depending on the communication delay of event 30; (4) ${\bf LOC}_{43}^C$ is at state 1 and disables event 43; (5) $tick$ will not be preempted, since no forcible event is enabled. If 30 is transmitted instantly, event 41 is enabled by ${\bf LOC}_{41}^C$ and the system will enter the automatic mode. If the transmission of 30 is delayed, only event $tick$ is enabled, and other events will not be enabled until the system enters the automatic mode. However, according to the transition diagram of ${\bf LOC}_{41}^C$ displayed in Fig.~\ref{fig:Local Controllers}, only after ${\bf LOC}_{41}^C$ has received the occurrence of event 30, will it enable event 41, and bring the system into the automatic mode.  Hence, the overall system behavior will not be affected even if the communication of event 30 is delayed. 

(2) Event 10 in $\Sigma_{V,com,O}$

\begin{figure}[!t]
\begin{center}
\includegraphics[scale = 0.5]{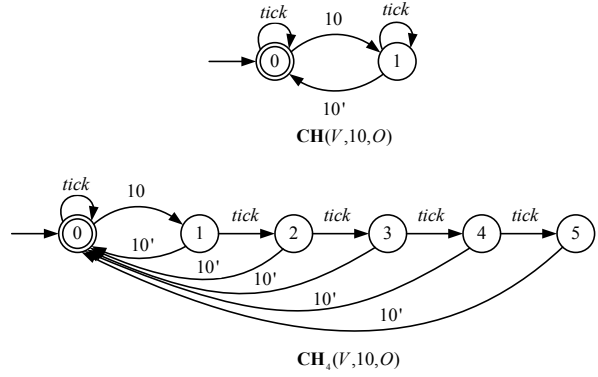}\\
\caption{Communication Channels ${\bf CH}(V,10,O)$ and ${\bf CH}_4(V,10,O)$}
\label{fig:COM10}
\end{center}
\end{figure}

Applying Algorithm 4, at Step 1 we verify by Algorithm 1 that $\bf SUP$ fails to be delay-robust with respect to the channel ${\bf CH}({V,10,O})$, as displayed in Fig.~\ref{fig:COM10}; then at Step 2, we check by Algorithm 3 that the maximal delay bound for event 10 is 4, i.e. $\bf SUP$ is bounded delay-robust with respect to the channel ${\bf CH}_4({V,10,O})$, as displayed in Fig.~\ref{fig:COM10}.

To illustrate that $\bf SUP$ is not delay-robust with respect to ${\bf CH}({V,10,O})$, but is bounded delay-robust with respect to ${\bf CH}_4({V,10,O})$, we consider the case that an increase in the voltage is reported (i.e. events 11 and 10 in $\bf VOLT$ have occurred sequentially). By inspection of the transition diagrams of the plant components shown in Fig.~\ref{fig:TDES Model}, the events that are eligible to occur are 11, 35, 41, 43, and $tick$. According to the transition diagrams of the local controllers and preemptors displayed in Fig.~\ref{fig:Local Controllers}, if $\bf OPTR$ knows the voltage increase before the fifth {\it tick} occurs, the tap-changer will generate a tap-up command and the operator can switch the system into manual mode; otherwise, the tap-changer will also generate a tap-up command, but the system cannot enter the manual mode. In terms of language, event 43 will be enabled after the event sequence $s := 11.10.tick.tick.tick.tick.tick.310.33$ (where event 310 is the signal event of 10), but is disabled after $s' := 11.10.tick. tick.tick.tick.tick.33$. When observing $s$ and $s'$ from the projection $P$ that erases the signal event 310, they cannot be distinguished. However, the system can enter the manual mode after the sequence $s$, but not after $s'$. In other words, the system can not `complete' the behavior of entering manual mode after $s'$, but this behavior can be finished in its delay-free counterpart $\bf SUP$. So, the observer property (\ref{eq:sub3:DefBDelayRobust}) required by bounded delay-robustness is violated when the delay bound $d$ exceeds 4 $tick$s, and we conclude that the maximal delay bound for event 10 is 4.

\begin{figure}[!t]
\begin{center}
\includegraphics[scale = 0.5]{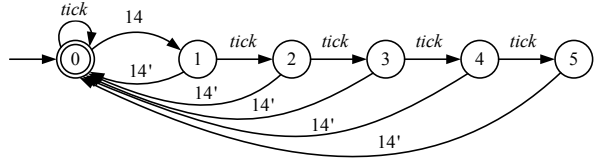}\\
\caption{Communication Channel ${\bf CH}_4(V,14,O)$}
\label{fig:COM14}
\end{center}
\end{figure}

Similarly, one can verify by Algorithm 4 that $\bf SUP$ is bounded delay-robust with respect to ${\bf CH}_4({V,14,O})$, as displayed in Fig.~\ref{fig:COM14}, and any other events except 10, 14 and 30 must be transmitted without delay.

(3) All communication events in (\ref{eq:ULTC:ComEvent})

Applying Algorithm 5 to each of the sets of communication events in (\ref{eq:ULTC:ComEvent}) in sequence, we obtain that $d_{max}'({T,30,O}) = \infty$, $d_{max}'({V,10,O}) = d_{max}'({V,14,O}) = 4$, and for the remaining events, $d_{max}' = 0$. In the following, we verify that if all the communication events are communicated within their corresponding delay bounds, the overall system behavior will still not be affected.

First, use ${\bf CH}({T,30,O})$, ${\bf CH}_4({V,10,O})$ and ${\bf CH}_4({V,14,O})$ to transmit events 30, 10 and 14 respectively. Second, connected by these channels, the overall system behavior is
\begin{align*}
{\bf SUP}_{com}' = &{\bf SUP}_V || {\bf SUP}_T || {\bf CH}_4({V,10,O}) || \\
& {\bf CH}_4({V,14,O})|| {\bf CH}({T,30,O}) || {\bf SUP}_O''')
\end{align*}
over the augmented alphabet $\{10,11,...,43,10',14',30'\}$, where ${\bf SUP}_O'''$ is obtained by replacing 10, 14, and 30 by $10'$, $14'$ and $30'$ respectively.
Third, one can verify that: (1) ${\bf SUP}_{com}'$ is correct and complete, and (2) ${\bf CH}({T,30,O})$, ${\bf CH}_4({V,10,O})$ and ${\bf CH}_4({V,14,O})$ will not cause uncontrollability with respect to the uncontrollable communication events. Finally, we conclude that the overall system behavior is still optimal and nonblocking.

\balance

\section{Conclusions} \label{Sec:Concl}


In this paper we have studied communication delays among local controllers  obtained by supervisor localization in TDES. First, we have identified properties of `timed delay-robustness' which guarantee that the specification of our delay-free distributed control continues to be enforced in the presence of (possibly unbounded) delay, and presented a polynomial verification algorithm to determine delay-robustness. Second, for those events that fail to be  delay-robust, we have proposed an algorithm to determine their maximal delay bound $d_{max}$ such that the system is $d_{max}$-bounded delay-robust. Finally, a ULTC example has exemplified these results, showing how to verify the delay-robustness, determine the maximal delay bound for bounded delay-robustness, and in addition, obtain a set of maximal delay bounds, one for each communication event, under the condition that the overall system behavior is still optimal and nonblocking.

With the definitions and tests reported here as basic tools, our future work will include the investigation of alternative more complex channel models and, of especial interest, global interconnection properties of a distributed system of TDES which may render delay-robustness more or less likely to be achieved.

\appendices
\section{Proof of Lemma~\ref{pro:BoundRelat}} \label{app:BoundDR}

To prove Lemma~\ref{pro:BoundRelat}, we need the following Lemmas~\ref{lem:SubsetRelation} and \ref{lem:TransformEquivalence}.
\begin{lem}\label{lem:SubsetRelation}
For any delay bound $d \geq 1$, there hold
\begin{align}
L({\bf SUP}) &\subseteq PL({\bf SUP}_d') \label{eq:sub1:SubsetRelation}\\
L_m({\bf SUP}) &\subseteq PL_m({\bf SUP}_d') \label{eq:sub2:SubsetRelation}
\end{align}
\end{lem}

\vspace{0.3cm} \noindent {\it Proof:}
Note that for different delay bounds $d$, the alphabets of ${\bf SUP}_d'$ and ${\bf CH}_d({j,\sigma,i})$ are $\Sigma' = \Sigma \cup \{\sigma'\}$ and $\{\sigma, tick,\sigma'\}$, respectively.
Here we only prove that $L({\bf SUP}) \subseteq PL({\bf SUP}_d')$; (\ref{eq:sub2:SubsetRelation}) can be proved in the same way by replacing $L$ by $L_m$ throughout.

Let $s \in L({\bf SUP})$; we must show that there exists a string $t\in L({\bf SUP}_d')$ such that $P(t) = s$. We first consider that only one instance of $\sigma$ appeared in $s$, and write $s = x_1\sigma x_2$ where $x_1,x_2$ are free of $\sigma$. By (\ref{eq:BNewModuBehavi}) and observing that ${\bf SUP}_i'$ is obtained by replacing each instance of $\sigma$ by $\sigma'$, we obtain that $t:=x_1\sigma\sigma'x_2 \in L({\bf SUP}_d')$. Furthermore,
$P(t) = s$. So, $L({\bf SUP}) \subseteq PL({\bf SUP}_d')$. This result can be easily extended to the general case that $s$ has multiple instances of $\sigma$, because $\sigma$ is transmitted by the channel model and the reoccurrence of $\sigma$ is permitted only when transmission of the previous $\sigma$ is completed. Namely, if $s = x_1\sigma x_2\sigma...,x_{k-1}\sigma x_k$, there exists a string $t = x_1\sigma\sigma' x_2\sigma\sigma'...,x_{k-1}\sigma\sigma' x_k$ such that $t \in L({\bf SUP}_d')$ and $P t = s$. Hence, we declare that $L({\bf SUP}) \subseteq PL({\bf SUP}_d')$.
\hfill $\blacksquare$


\begin{lem}\label{lem:TransformEquivalence}
Let $t = x_1\sigma x_2 x_3\sigma'x_4\in L_m({\bf SUP}_d')$ where $x_1,x_2,x_3~\text{and}~x_4$ are strings free of $\sigma$ and $\sigma'$, i.e. $x_1,x_2,x_3,x_4 \in (\Sigma - \{\sigma\})^*$. Then $t' := x_1\sigma x_{2} \sigma'x_{3}x_4 \in L_m({\bf SUP}_d')$.
\end{lem}


\vspace{0.3cm} \noindent {\it Proof of Lemma~\ref{lem:TransformEquivalence}:}
Recall that ${\bf SUP}_i'$ is ${\bf SUP}_i$ with transitions labeled $\sigma$ relabeled $\sigma'$. By definition of synchronous product, $x_2$, $x_3$ and $\sigma'$ can be re-ordered without affecting the membership of $t$ in $L_m({\bf SUP}_d')$, namely the strings $t'$ formed from $t$ by the successive replacement
\begin{align*}
x_1\sigma x_2 x_3\sigma'x_4 &\rightarrow x_1\sigma \sigma'x_2 x_3 x_4\\
                              &\rightarrow x_1\sigma x_2 \sigma' x_3x_4
\end{align*}
will belong to $L_m({\bf SUP}_d')$ as well. In other words, if the transmission of $\sigma$ is completed in a shorter time (the number of {\it tick}s in $x_2$ will be smaller than that in $x_2x_3$), the behavior is still legal.
\hfill $\blacksquare$


\vspace{0.3cm} \noindent {\it Proof of Lemma \ref{pro:BoundRelat}:}
We prove Lemma~\ref{pro:BoundRelat} by contraposition, i.e. if ${\bf
SUP}$ is $(d+1)$-bounded delay-robust, then it is also
$d$-bounded delay-robust. To that end, we must verify (\ref{eq:sub1:DefBDelayRobust})-(\ref{eq:BControllabilityTest}).
\vspace{0.1in}

(1) For (\ref{eq:sub1:DefBDelayRobust}), we prove that $PL({{\bf SUP}_{d}}') \supseteq L({\bf SUP})$ and $PL({{\bf SUP}_{d}}') \subseteq L({\bf SUP})$ in sequence. $PL({{\bf SUP}_{d}}') \supseteq L({\bf SUP})$ is obtained from Lemma~\ref{lem:SubsetRelation} immediately. By inspection
of the transition diagram of ${\bf CH}_d({j,\sigma,i})$ in Fig.~\ref{fig:BoundComChannel}, we get that $L({\bf CH}_{d}({j,\sigma,i})) \subseteq L({\bf CH}_{d+1}({j,\sigma,i}))$. So according to (\ref{eq:BNewModuBehavi}),
\begin{align} \label{eq:SubseteqBoundBehavi}
L({\bf SUP}_{d}') \subseteq L({\bf SUP}_{d+1}').
\end{align}
Since $\bf SUP$ is $(d+1)-$bounded delay-robust, $PL({\bf SUP}_{d+1}') \subseteq L({\bf SUP})$. Hence, $PL({{\bf SUP}_{d}}') \subseteq L({\bf SUP})$.
\vspace{0.1in}

(2) Condition (\ref{eq:sub2:DefBDelayRobust}) can be confirmed from the proof of
(\ref{eq:sub1:DefBDelayRobust}) by replacing $L$ by $L_m$ throughout.
\vspace{0.1in}

(3) For (\ref{eq:sub3:DefBDelayRobust}), assume that
$s \in L({\bf SUP}_{d}')$ and $(Ps)w \in L_m({\bf SUP})$; we must show that there exists a string $v \in \Sigma'^*$ such that $Pv = w $ and $sv \in L_m({\bf SUP}_{d}')$.

By (\ref{eq:SubseteqBoundBehavi}), we have $s \in L({\bf SUP}_{d+1}')$. Since $\bf SUP$
is $(d+1)-$bounded delay-robust, there exists a string $u \in \Sigma'^*$ such that
$Pu = w$ and $su \in L_m({\bf SUP}_d')$. Here we consider the case that only one instance
of $\sigma$ exists in $su$; the general cases can be confirmed similarly (since the
transmission of multiple instances of $\sigma$ does not result in mutual interference).
In the following, we prove (\ref{eq:sub3:DefBDelayRobust}) from these three cases:
(i) $su = s_1\sigma s_2\sigma's_3u_1u_2$, (2) $su = s_1 \sigma s_2 u_1\sigma' u_2$,
and (iii) $s_1s_2 u_1 \sigma u_2 \sigma' u_3$, where $s_1, s_2, s_3, u_1, u_2, u_3$ are free of
$\sigma$ and $\sigma'$.

(\rmnum{1}) $su = s_1\sigma s_2\sigma's_3 u_1u_2$. By (\ref{eq:BNewModuBehavi}), we have $su \in L_m({\bf NSUP})$. Similarly, since $s \in L({\bf SUP}_d')$, $s \in P_{ch}^{-1}L({\bf CH}_{d}({j,\sigma,i}))$. Further, $s = s_1\sigma s_2\sigma' s_3$, which means that after string $s$, $\sigma'$ has reset the channel ${\bf CH}_{d}({j,\sigma,i})$. Thus $s \in P_{ch}^{-1}L_m({\bf CH}_{d-1}({j,\sigma,i}))$.  On the other hand, because $u$ is free of $\sigma$, $su \in P_{ch}^{-1}L_m({\bf CH}_{d}({j,\sigma,i}))$. Hence, $su \in L_m({\bf SUP}_{d}')$. Define $v = u$;
then $Pv = Pu = w $ and $sv \in L_m({\bf SUP}_{d}')$, as required by (\ref{eq:sub3:DefBDelayRobust}).

(\rmnum{2}) $su = s_1 \sigma s_2 u_1\sigma' u_2$. By Lemma \ref{lem:TransformEquivalence},, it results from $su \in L_m({\bf SUP}_d')$ that $s_1\sigma s_2\sigma'u_1u_2 \in L_m({\bf SUP}_d')$. The rest is similar to case (1); in this case, $v = \sigma'u_1u_2$.

(\rmnum{3}) $su = s_1s_2 u_1 \sigma u_2 \sigma' u_3$. By Lemma \ref{lem:TransformEquivalence}, we have $s_1s_2 u_1 \sigma \sigma' u_2 u_3\in L_m({\bf SUP}_d')$. Also, the rest is similar to case (1); in this case, $v = u_1\sigma\sigma'u_2u_3$.
\vspace{0.1in}

(4) Let $s \in P_{ch}^{-1}L({\bf CH}_{d}({j,\sigma,i}))$ and $s\sigma \in L({\bf NSUP})$; we show that $s \sigma \in P_{ch}^{-1}L({\bf CH}_{d}({j,\sigma,i}))$ by contraposition. Assume that $s \sigma \notin P_{ch}^{-1}L({\bf CH}_{d}({j,\sigma,i}))$. Write ${\bf CH}_d({j,\sigma,i}) = (C_d, \Sigma_{ch},\tau_d,c_{d,0}, \{c_{d,0}\})$ where $\Sigma_{ch} = \{\sigma,tick,\sigma'\}$. We claim that $\tau_{d}(c_{d,0}, P_{ch}s) \neq c_{d,0}$; otherwise, $\sigma$ is defined at state $\tau_{d}(c_{d,0}, P_{ch}s)$ and $s\sigma \in P_{ch}^{-1}L({\bf CH}_{d}({j,\sigma,i}))$. By inspection of the transition diagrams of ${\bf CH}_{d}({j,\sigma,i})$ and ${\bf CH}_{d + 1}({j,\sigma,i})$, it results from $\tau_{d}(c_{d,0}, P_{ch}s) \neq c_{d,0}$, that $\tau_{d+1}(c_{d+1,0}, P_{ch}s) \neq c_{d+1,0}$. Hence, $s\sigma \notin P_{ch}^{-1}L({\bf CH}_{d+1}({j,\sigma,i}))$, in contradiction to the fact that $\bf SUP$ is $(d+1)-$bounded delay-robust.
\hfill $\blacksquare$

\section{Proof of Lemma~\ref{pro:Alg3FiniteTerminate}} \label{app:SpecDR}



Since $\bf SUP$ is not delay-robust wrt. ${\bf CH}(j,\sigma,i)$, by Definition~\ref{defn:unboundDelayRobust}, one of the conditions (\ref{eq:sub1:DefDelayRobust})-(\ref{eq:ControllabilityTest}) is violated.
In the following, we prove that in each case, $d_{max} \leq 2^m *m$, where
$m$ is the states number of ${\bf SUP}'$ in (\ref{eq:NewModuBehavi}).

\begin{figure}[!t]
\centering
    \includegraphics[scale = 0.6]{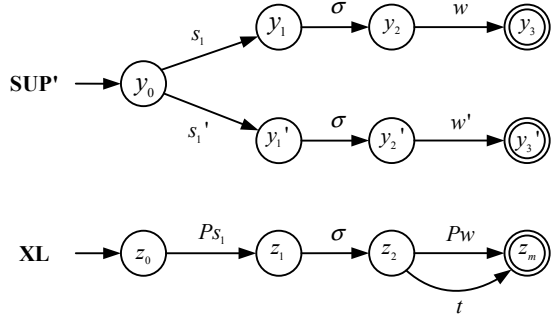}
\caption{$Ps_1 = Ps_1'$, $Pw' = t$ and $t$ is a simple string.}
\label{fig:XL}
\end{figure}

(1) Condition (\ref{eq:sub1:DefDelayRobust}) is violated. Since that $L({\bf SUP})\subseteq PL({\bf SUP}')$ always holds (similar to Lemma~\ref{lem:SubsetRelation}), we have $PL({\bf SUP}') \nsubseteq L({\bf SUP})$. So, there exists at least one string $s \in \Sigma'^*$ such that $s \in L({\bf SUP}')$, but $Ps \notin L({\bf SUP})$. We claim that $s$ can be written as $s_1\sigma w$ where $s_1, w \in \Sigma'^*$; otherwise, $s$ does not contain any $\sigma$, and it follows from the construction of ${\bf SUP}'$ that $Ps \in L({\bf SUP})$, a contradiction. As illustrated in Fig.~\ref{fig:XL}, we prove in the following that there exist strings $s_1' \in L({\bf SUP}')$ and $w' \in \Sigma'^*$ such that $\#tick(w') \leq 2^m*m$ (where $\#tick(w')$ represents the number of events $tick$ appearing in string $w'$), $s'\sigma w' \in L({\bf SUP}')$, but $P(s'\sigma w') \notin L({\bf SUP})$, from which we can conclude: to prevent the occurrence of string $s_1'\sigma w'$, the maximal communication delay of $\sigma$ must be less than $\#tick(w') \leq 2^m*m$, i.e. $d_{max} \leq 2^m *(m' + 1)$.

By $s_1\sigma w \in L({\bf SUP}')$ and $P(s_1\sigma w) \notin L({\bf SUP})$, we have $P(s_1\sigma w) \in PL({\bf SUP}')\cap (\Sigma^* - L({\bf SUP}))$. To identify such strings, we build an TDES ${\bf XL} = (Z,\Sigma,\zeta,z_0,Z_m)$ such that
\[L_m({\bf XL}) = PL({\bf SUP}')\cap (\Sigma^* - L({\bf SUP}))\]
and
\[L({\bf XL}) = PL({\bf SUP}'),\]
i.e., $P(s_1\sigma) \in L({\bf XL})$, and $P(s_1\sigma w) \in L_m({\bf XL})$.

First, we build $\bf XA$ such that $L_m({\bf XA}) = PL({\bf SUP}')$ and $L({\bf XA}) = L_m({\bf XA})$ by the following two steps: (\rmnum{1}) construct ${\bf PSUP}'$ by applying the subset construction algorithm on ${\bf SUP}'$ with natural projection $P$, and (\rmnum{2}) obtain ${\bf XA}$ by marking all states of ${\bf PSUP}'$.
Second, we build $\bf XB$ such that $L_m({\bf XB}) = \Sigma^* - L({\bf SUP})$
and $L({\bf XB}) = \Sigma^*$ by first adjoining a (non-marker) dump state $\hat{q}$ to the state set of $\bf SUP$ and transitions $(q,\sigma,\hat{q})$ for each state $q$ of $\bf SUP$ if $\sigma \in \Sigma$ is not defined at $q$ (i.e. $L({\bf XB}) = \Sigma^*$), and secondly setting $\hat{q}$ to be the only marker state.
Third, let ${\bf XL} = {\bf XA} || {\bf XB}$; then $L_m({\bf XL}) = PL({\bf SUP}')\cap (\Sigma^* - L({\bf SUP}))$, $L({\bf XL}) = PL({\bf SUP}')$.
The state size $|Z| \leq 2^m*(m'+1)$, since ${\bf XA}$ has at most $2^m$ states (due to the subset construction algorithm), and ${\bf XB}$ has $m' + 1$ states .

Finally, by $P(s_1\sigma) \in PL({\bf SUP}') = L({\bf XL})$, there exists a state $z_2 \in Z$ such that $z_2 = \zeta(z_0, P(s\sigma))$; by $P(s_1\sigma w) \in L_m({\bf XL})$, there exists a marker state $z_m \in Z_m$ such that $z_m = \zeta(z_0, P(s_1\sigma w)) = \zeta(z_2, P(w))$. So, there exists at least a {\it simple string}\footnote{The concept `simple string' is derived from the `simple path' in graphic theory, where a path is called {\it simple}
if no vertex is traversed more than once\cite{Danielson:1968}. Here string $t$ is called {\it simple} if no state is traversed more than once.} $t \in \Sigma^*$ joining $z_2$ and $z_m$ such that $z_m = \zeta(z_2,t)$, and thus $P(s_1\sigma) t \in L_m({\bf XL})$. It follows that $(Ps_1)\sigma t \in PL({\bf SUP}')\cap (\Sigma^* - L({\bf SUP}))$. So, there exist strings $s_1', w' \in \Sigma'^*$ such that $Ps_1' = Ps_1$, $Pw' = t$, $s_1'\sigma w' \in L({\bf SUP}')$, and $P(s_1'\sigma w') \notin L({\bf SUP})$, namely the occurrence of $w'$ after $s_1'\sigma$ violates condition (\ref{eq:sub1:DefDelayRobust}). Since $t$ is simple, we have $\#tick(t) \leq |Z| \leq 2^m*(m'+1)$. By $Pw' = t$, we have $\#tick(w') = \#tick(t) \leq 2^m*(m'+1)$. Furthermore, since ${\bf SUP}'$ represents the system behavior with communication delay, we always have $m' + 1 \leq m$. So $\#tick(w') \leq 2^m *m$, as required.

(2) Condition (\ref{eq:sub2:DefDelayRobust}) is violated. $d_{max} \leq m*2^{m}$ can be confirmed similar to case (1).

\begin{figure}[!t]
\centering
    \includegraphics[scale = 0.6]{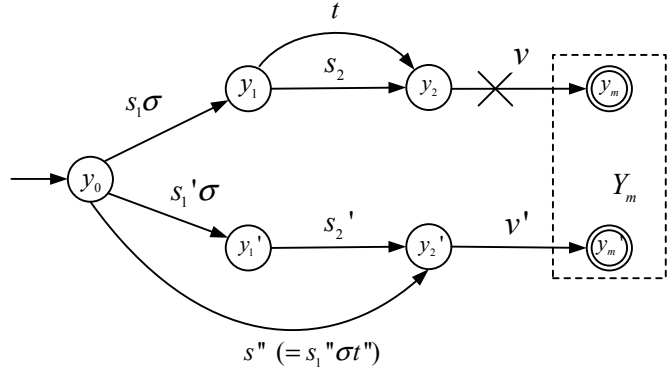}
\caption{$P$-Normality of ${\bf SUP}'$}
\label{fig:PNormal}
\end{figure}

(3) Condition (\ref{eq:sub3:DefDelayRobust}) is violated. Since delay-robustness of $\bf SUP$ is violated by the communication delay of $\sigma$, there must exist strings $s_1$, $s_2$, and $w$, such that $s_1\sigma s_2 \in L({\bf SUP}')$ and $P(s_1\sigma s_2) w \in L_m({\bf SUP})$, but no string $v$ satisfies that $Pv = w$ and $s_1\sigma s_2 v \in L_m({\bf SUP}')$. As illustrated in Fig.\ref{fig:PNormal}, we prove in the following that the condition (\ref{eq:sub3:DefDelayRobust}) is also violated by the string pair $s_1 \sigma t$ and $s_1''\sigma t''$ where $\#tick(t) \leq 2^m*m$ and $\#tick(t'') \leq 2^m*m$, from which we conclude: to prevent the occurrences of the strings $s_1 \sigma t$ and $s_1''\sigma t''$, the communication delay of $\sigma$ must be less than $min(\#tick(t),\#tick(t'')) \leq 2^m*m$ , i.e. $d_{max}\leq m*2^{m}$.

To that end, we need the concept `normal automaton'\cite{TakaiUshio:2003}. For ${\bf SUP}' = (Y,\Sigma',\eta,y_0,Y_m)$, we say that ${\bf SUP}'$ is $P$-normal if
\begin{align} \label{eq:normal}
(\forall s, t \in L({\bf SUP}')) R(s) \neq R(t) \Rightarrow R(s) \cap R(t) = \emptyset
\end{align}
where $R(s):= \{y \in Y|y = \eta(y_0,s'), Ps = Ps'\}$. In case ${\bf SUP}'$ is not $P$-normal, replace ${\bf SUP}'$ by ${\bf SUP}' || {\bf PSUP}'$ where ${\bf PSUP}'$ is a deterministic generator over $\Sigma$ obtained by the subset construction. ${\bf SUP}' || {\bf PSUP}'$ is always $P$-normal, and $L({\bf SUP}') = L({\bf SUP}' || {\bf PSUP}')$ and $L_m({\bf SUP}') = L_m({\bf SUP}' || {\bf PSUP}')$. The state size of the new ${\bf SUP}'$ is at most $m*2^m$.

By $P(s_1\sigma s_2) w \in L_m({\bf SUP}) \subseteq PL_m({\bf SUP}')$, there must exist strings $s_1'$, $s_2'$, and $v'$ such that $Ps_1' = Ps_1$, $Ps_2' = Ps_2$, $Pv' = w$, and $s_1'\sigma s_2' v' \in L_m({\bf SUP}')$, as displayed in Fig.~\ref{fig:PNormal}. Let $y_1 = \eta(y_0, s_1\sigma)$, $y_2 = \eta(y_1, s_2)$, $y_1' = \eta(y_0, s_1'\sigma)$, and
$y_2' = \eta(y_1', s_2')$. Joining $y_1$ and $y_2$, there must exist a simple string $t$ such that $y_2 = \eta(y_1, t)$. So, $R(s_1\sigma s_2) \cap R(s_1\sigma t) = y_2$. By $P$-normality of ${\bf SUP}'$, there must exist a string $s'' \in L({\bf SUP}')$ such that $y_2' = \eta(y_0, s'')$, $P(s_1\sigma t) = P(s'')$, and $y_2' \in R(s_1 \sigma t)$. So string $s''$ can be written as $s_1''\sigma t''$ where $Ps_1'' = Ps_1$ and $Pt'' = Pt$, and the condition (\ref{eq:sub3:DefDelayRobust}) is also violated by the string pair $s_1 \sigma t$ and $s_1''\sigma t''$. Because $t$ is simple, $\#tick(t) \leq m$, where $m$ is the state size of $P$-normal form of ${\bf SUP}'$. So, when ${\bf SUP}'$ is not $P$-normal, $\#tick(t) \leq m*2^m$. In addition, since $Pt'' = Pt$, $\#tick(t'') = \#tick(t) \leq m*2^m$, as required.

(4) Condition (\ref{eq:ControllabilityTest}) is violated. In this case, assume that $\sigma$ is blocked at state $y$ of ${\bf SUP}'$, and the last occurrence of $\sigma$ occurs at state $y'$ of ${\bf SUP}'$. From $y'$ to $y$, there must exist a simple string $t$. We claim that the maximal communication delay of $\sigma$ must be less that $\#tick(t)$; otherwise, the system will arrive at state $y$ by string $t$. Hence $d_{max}\leq \#t(tick) \leq m$.

Finally, by comparing $d_{max}$ in the above four cases, we conclude that if $\bf SUP$ is not delay-robust with respect to ${\bf CH}(j,\sigma,i)$, $d_{max}\leq m*2^{m}$.



\bibliographystyle{IEEEtran}
\bibliography{SCDES_Ref}

\end{document}